\documentclass[%
 reprint,
 superscriptaddress,
preprintnumbers,
nofootinbib,
 amsmath,amssymb,
 aps,
 prd,
]{revtex4-1}

\usepackage[percent]{overpic}
\usepackage{tikz}
\usepackage{graphicx}
\usepackage{dcolumn}
\usepackage{bm}
\usepackage{hyperref}
\usepackage{color}
\usepackage{array} 
\usepackage{float} 

\definecolor{airforceblue}{rgb}{0.36, 0.54, 0.66}
\definecolor{darkmidnightblue}{rgb}{0.0, 0.2, 0.4}

\definecolor{royalazure}{rgb}{0.0, 0.22, 0.66}

\usepackage{multirow, array}

\newcommand{\dd}{\textnormal{d}}

\hypersetup{
    colorlinks,
    linktoc=page,
    citecolor = royalazure,
}
\usepackage{aas_macros}

\renewcommand{\vec}{\bm}

\def\fun#1#2{\lower3.6pt\vbox{\baselineskip0pt\lineskip.9pt
        \ialign{$\mathsurround=0pt#1\hfill##\hfil$\crcr#2\crcr\sim\crcr}}}

\newcommand{\beq}{\begin{equation}}
\newcommand{\eeq}{\end{equation}}
\newcommand{\beqa}{\begin{eqnarray}}
\newcommand{\eeqa}{\end{eqnarray}}
\newcommand{\be}{\begin{equation}}
\newcommand{\ee}{\end{equation}}
\newcommand{\bea}{\begin{eqnarray}}
\newcommand{\eea}{\end{eqnarray}}
\newcommand{\nn}{\nonumber}

\newcommand{\asz}{\ensuremath{{\zeta_0}}}
\newcommand{\bsz}{\ensuremath{{\zeta_\mathrm{M}}}}
\newcommand{\csz}{\ensuremath{{\zeta_\mathrm{z}}}}
\newcommand{\dsz}{\ensuremath{{\sigma_{\ln\zeta}}}}

\newcommand{\alambda}{\ensuremath{{\lambda_0}}}
\newcommand{\blambda}{\ensuremath{{\lambda_\mathrm{M}}}}
\newcommand{\clambda}{\ensuremath{{\lambda_\mathrm{z}}}}
\newcommand{\dlambda}{\ensuremath{{\sigma_{\ln\lambda}}}}
\newcommand{\sigmalnlambda}{\ensuremath{{\sigma_{\ln\lambda}}}}
\newcommand{\aWL}{\ensuremath{{\ln M_{\mathrm{WL}_0}}}}
\newcommand{\bWL}{\ensuremath{{M_{\mathrm{WL}_\mathrm{M}}}}}
\newcommand{\asigmaWL}{\ensuremath{{\ln\sigma^2_{\ln\mathrm{WL}_0}}}}
\newcommand{\bsigmaWL}{\ensuremath{{\sigma_{\ln\mathrm{WL}_\mathrm{M}}}}}

\newcommand{\Mwl}{\ensuremath{M_\mathrm{WL}}}
 
\newcommand{\Msun}{\ensuremath{\mathrm{M}_\odot}}

\definecolor{pygreen}{RGB}{44, 160, 44}

\usepackage[normalem]{ulem}

\begin{document}

\preprint{TUM-HEP 1490/23}

\title{Probing interacting dark sector models with future weak lensing-informed galaxy cluster abundance constraints from SPT-3G and CMB-S4}

\author{Asmaa Mazoun}
\email{asmaa.mazoun@tum.de}
\affiliation{\small Physik Department T31, Technische Universit\"at M\"unchen,
James-Franck-Stra\ss e 1, D-85748 Garching, Germany}
\affiliation{\small University Observatory, Faculty of Physics, Ludwig-Maximilians-Universität, Scheinerstr. 1, D-81679 München, Germany}
\affiliation{Excellence Cluster ORIGINS, Boltzmannstr. 2, D-85748 Garching, Germany}
\author{Sebastian Bocquet}
\affiliation{\small University Observatory, Faculty of Physics, Ludwig-Maximilians-Universität, Scheinerstr. 1, D-81679 München, Germany}

\author{Mathias Garny}
\affiliation{\small Physik Department T31, Technische Universit\"at M\"unchen,
James-Franck-Stra\ss e 1, D-85748 Garching, Germany}

\author{Joseph~J.~Mohr}
\affiliation{\small University Observatory, Faculty of Physics, Ludwig-Maximilians-Universität, Scheinerstr. 1, D-81679 München, Germany}
\affiliation{Max Planck Institute for Extraterrestrial Physics, Gießenbachstr.~1, D-85748 Garching, Germany}

\author{Henrique Rubira}
\affiliation{\small Physik Department T31, Technische Universit\"at M\"unchen,
James-Franck-Stra\ss e 1, D-85748 Garching, Germany}
\affiliation{Excellence Cluster ORIGINS, Boltzmannstr. 2, D-85748 Garching, Germany}

\author{Sophie M. L. Vogt}
\affiliation{\small University Observatory, Faculty of Physics, Ludwig-Maximilians-Universität, Scheinerstr. 1, D-81679 München, Germany}
\affiliation{Excellence Cluster ORIGINS, Boltzmannstr. 2, D-85748 Garching, Germany}
\affiliation{Max Planck Institute for Astrophysics, Karl-Schwarzschild-Str.~1, D-85748 Garching, Germany}


\begin{abstract} 

We forecast the sensitivity of ongoing and future galaxy cluster abundance measurements to detect deviations from the cold dark matter (CDM) paradigm. Concretely, we consider a class of dark sector models that feature an interaction between dark matter and a dark radiation species (IDM--DR). This setup can be naturally realized by a non-Abelian gauge symmetry and has the potential to explain $S_8$ tensions arising within $\Lambda$CDM.
We create mock catalogs of the ongoing SPT-3G as well as the future CMB-S4 surveys of galaxy clusters selected via the thermal Sunyaev-Zeldovich effect (tSZE). Both datasets are complemented with cluster mass calibration from next-generation weak gravitational lensing data (ngWL) like those expected from the Euclid mission and the Vera C. Rubin Observatory. We consider an IDM--DR scenario with parameters chosen to be in agreement with Planck 2018 data and that also leads to a low value of $S_8$ as indicated by some local structure formation analyses. 
Accounting for systematic and stochastic uncertainties in the mass determination and the cluster tSZE selection, we find that both SPT-3G$\times$ngWL and CMB-S4$\times$ngWL cluster data will be able to discriminate this IDM--DR model from $\Lambda$CDM, and thus test whether dark matter -- dark radiation interactions are responsible for lowering $S_8$. Assuming IDM--DR, we forecast that the temperature of the dark radiation 
can be determined to about 40\% (10\%) with SPT-3G$\times$ngWL (CMB-S4$\times$ngWL), considering 68\% credibility, while $S_8$ can be recovered with percent-level accuracy. Furthermore, we show that IDM--DR can be discriminated from massive neutrinos, and that cluster counts will be able to constrain the dark radiation temperature to be below $\sim 10\%$ (at 95\% credibility) of the cosmic microwave background temperature if the true cosmological model is $\Lambda$CDM.

\end{abstract}

\maketitle


\section{Introduction}
\label{sec:introduction}

Dark matter (DM) plays a key role in the formation of structure during cosmological evolution. Although it represents most of the matter component in the Universe, the DM particle has still not been detected directly, indirectly or through its production in accelerators like the Large Hadron Collider. 
Nevertheless, cosmological probes at different cosmological times and scales are sensitive to fundamental properties of dark matter.
On large ($\gtrsim$ Mpc) scales these include the cosmic microwave background anisotropies (CMB)~\cite{Planck:2018vyg}, baryon acoustic oscillations (BAO)~\cite{eBOSS:2020yzd}, galaxy cluster counts~\cite{SPT:2018njh,Chiu:2022qgb},  galaxy clustering~\cite{BOSS:2016wmc, Reid:2015gra}, and weak gravitational lensing \cite{Heymans:2020gsg, DES:2021wwk, Kilo-DegreeSurvey:2023gfr,DES+Kids23}.  
Next-generation CMB and large-scale structure (LSS) observatories, such as CMB-S4 \cite{CMB-S4:2016ple,Abazajian:2019eic}, Euclid \cite{Laureijs2011, EuclidTheoryWorkingGroup:2012gxx,Euclid:2021icp}, and the Vera C. Rubin Observatory \cite{LSST:2008ijt} will significantly enhance our level of understanding of DM by allowing for precision tests of cosmological evolution on Mpc scales.

Within the standard cold dark matter (CDM) paradigm, DM  is assumed to be cold, non-relativistic and only gravitationally interacting. This simple picture has been successful in explaining most of the observed astrophysical and cosmological phenomena. Nevertheless, beyond-CDM models may provide solutions to possible open questions in cosmology and astrophysics, such as the $H_0$ tension~\cite{Verde:2019ivm, Riess:2019qba}, the $S_8$ tension~\cite{Nunes:2021ipq, Amon:2022azi} and small-scale problems~\cite{Hui:2016ltb, Bullock:2017xww, Tulin:2017ara}. Moreover, they can provide a window to explore other possible properties of dark matter, e.g., DM self-interaction~\cite{Egana-Ugrinovic:2021gnu}, long-range interactions~\cite{Bottaro:2023wkd} or its decay to other particles~\cite{Simon:2022ftd, Fuss:2022zyt}.
In analogy to the Standard Model of particle physics, recent works draw the attention to the possibility of having a secluded {\it dark sector} which contains an interacting dark matter (IDM) and a new relativistic component named dark radiation (DR)~\cite{Buen-Abad:2015ova,Lesgourgues:2015wza,Cyr-Racine:2015ihg,Buen-Abad:2017gxg, Archidiacono:2019wdp, Becker:2020hzj, Rubira:2022xhb}. The interacting dark matter --  dark radiation (IDM--DR) model is well motivated within particle physics~\cite{Buen-Abad:2015ova,Rubira:2022xhb} and can lead to a lower value of $S_8$ while remaining in agreement with CMB data, potentially solving the aforementioned $S_8$ tension. We note that an extension of this model known as interacting stepped dark radiation has been introduced and studied and shown to be relevant to both $H_0$ and $S_8$ tensions~\cite{Aloni:2021eaq, Joseph:2022jsf, Schoneberg:2023rnx}. In this work, we restrict ourselves to the simpler version of the IDM--DR model, although the methodology can in principle be extended to more complicated dark sector models in future work.  

The quantity $S_8=\sigma_8 \sqrt{\Omega_{\rm m}/0.3}$ is a combination of the matter density parameter $\Omega_{\rm m}$ and the fluctuation amplitude in the linear matter density at scales of 8 $h^{-1}$Mpc, $\sigma_8$ that is well constrained by local structure formation analyses. Its value measured via weak lensing by KiDS~\cite{Heymans:2020gsg}, DES~\cite{DES:2021wwk} and HSC~\cite{HSC:2018mrq}, galaxy cluster counts by SPT~\cite{SPT:2018njh} and eROSITA~\cite{Chiu:2022qgb}, and possibly also Lyman-$\alpha$ forest data~\cite{Fernandez:2023grg} by eBOSS~\cite{Chabanier:2018rga} as well as the cross correlation of electron pressure fluctuations as inferred from the thermal Sunyaev-Zeldovich effect (tSZE) on the CMB with galaxy clustering~\cite{Chen:2023hyg} is lower than the one inferred from Planck 2018 CMB anisotropies~\cite{Planck:2018vyg} when assuming the standard $\Lambda$CDM model, with a tension of $\sim 2-3 \,\sigma$, depending on the dataset. An analysis within the IDM--DR model that takes CMB anisotropies (Planck 2015) and BAO datasets into account showed a degeneracy between $S_8$ and the dark radiation temperature (parameterized by its ratio $\xi_{\rm DR}$ to the CMB temperature, see Eq.~\ref{eq:xidef}), and found an allowed direction in parameter space accommodating for a lower value of $S_8$~\cite{Archidiacono:2019wdp}. Subsequently, it was shown that this preference is enhanced when adding either redshift-space distortion (RSD) or the full shape (FS) information from BOSS galaxy clustering data~\cite{BOSS:2016wmc} combined with Planck 2018, allowing for values of $S_8$ that are nominally within $\sim 1\sigma$ of those extracted from weak lensing measurements for $\xi_{\rm DR}\sim {\cal O}(0.1)$~\cite{Rubira:2022xhb}. Nevertheless, these analyses resulted in only upper bounds on $\xi_{\rm DR}$, rather than a definitive discrimination between the $\Lambda$CDM and IDM--DR model.

We investigate for the first time whether galaxy cluster abundance analyses with weak-lensing mass calibration are sensitive to dark matter -- dark radiation interactions, can discriminate CDM from IDM--DR at the level relevant for addressing the $S_8$ tension, and can enable tighter constraints on IDM--DR model parameters. A galaxy cluster is the observable signature of a dark matter halo that has formed from the highest peaks in the matter density field on large scales. This makes the abundance of  clusters particularly sensitive to the density of matter and the amplitude of matter density fluctuations, allowing one to constrain $\Omega_{\rm m}$ and $\sigma_8$ as well as the properties of 
dark energy, which impact the structure formation history~\cite{Haiman:2000bw}.
Specifically, in this work, we consider galaxy cluster samples detected in mm-wave data through the tSZE, where CMB photons scatter with electrons in the hot, intracluster medium. 

Optical and near-infrared (NIR) followup of mm-wave mapping experiments with a clean and well-understood selection, have been employed to create samples of tSZE selected galaxy clusters, such as SPT \cite{Staniszewski09,Bleem20ApJS..247...25B,Klein23arXiv230909908K,Bleem23arXiv231107512B}, ACT \cite{ACT:2020lcv}, and Planck \cite{Planck:2015koh,Hernandez-Lang23MNRAS.525...24H}. 
Crucial to the cosmological analyses of these samples is the use of weak gravitational lensing to constrain the cluster masses 
\cite{Applegate_2014, Hoekstra:2015gda,SPT:2018njh} as well as the careful accounting for systematic and stochastic uncertainties in constraining the so-called observable--mass relations \cite{Grandis:2021aad,Bocquet23arXiv231012213B}.
We base our work on the analysis framework adopted by SPT, which has enabled competitive constraints on $\Lambda$CDM and $w$CDM cosmologies using galaxy clusters selected in SPT-SZ data \citep{SPT:2018njh} and in the combination of SPT-SZ and SPTpol data \citep{Bocquet24}.
An analysis of IDM--DR models using the latest SPT-selected cluster sample together with DES weak gravitational lensing data is underway.  
Building upon this experience,
in this forecast paper, we characterize the promise of future experiments by mocking up tSZE selected cluster catalogs for two different surveys: SPT-3G~\cite{Carlstrom11PASP..123..568C,Benson14SPIE.9153E..1PB} and CMB-S4~\cite{CMB-S4:2016ple}. For both cluster samples, the mass calibration relies upon weak-lensing data of the quality expected from the ongoing Euclid mission or the future Rubin Observatory.  Hereafter, we refer to these next-generation weak gravitational lensing surveys as ngWL.

The article begins with an introduction to the interacting dark matter--dark radiation (IDM--DR) model and its main parameters, and a discussion of its effect on the matter power spectrum and the halo mass function (HMF; Sec~\ref{sec:IDM}). In Sec~\ref{sec:data} we describe the forecast data used in the analysis with a brief explanation of the method used to generate mocks and perform the mass calibration. An overview of the analysis appears in Sec~\ref{sec:analysis}, and in Sec~\ref{sec:results} we present the results obtained for constraining the model parameters and testing whether IDM--DR is responsible for lowering $S_8$, including a discussion about adding the mass of neutrinos as a free parameter and its interplay with the sensitivity of cluster counts to interacting dark sectors.

\section{Interacting Dark Matter-Dark Radiation (IDM--DR) model}
\label{sec:IDM}
In this work we go beyond the CDM paradigm by studying non-gravitational dark matter interactions which have an impact on structure formation on large ($\gtrsim$ Mpc) scales. In analogy to the visible sector where electrons and photons interact, we consider a similar scattering in the dark sector between the fraction of the dark matter that is interacting (IDM) and a new relativistic component, called dark radiation (DR).
Depending on the particle physics model considered, DR can also undergo self-interaction. The remainder of the DM behaves as the usual cold dark matter (CDM). For the IDM--DR model considered here, this interaction is relevant over an extended period of time, having more impact before matter-radiation equality without substantially affecting structure formation afterwards. From a particle physics perspective there could be different realizations of such an interaction in the dark sector, yet the details of these models could be translated to effective parameters for the cosmological analysis. This mapping is done in the framework of Effective THeory Of Structure formation (ETHOS ~\cite{Cyr-Racine:2015ihg}).

\subsection{Formalism}

In this section we summarize the framework describing the evolution of perturbations for IDM--DR interaction (see \cite{Cyr-Racine:2015ihg} for a more complete description). Besides the equations describing $\Lambda$CDM components, extra equations are included for IDM and DR. For non-relativistic IDM only the two first moments are needed; the density contrast $\delta_{\rm IDM}$ and the velocity divergence $\theta_{\rm IDM}$. Because the scattering is elastic, the number of particles is conserved and so the continuity equation of the density remains the same as for CDM. On the other hand, the IDM--DR interaction leads to a momentum exchange and a drag force term in the Euler equation,
\begin{eqnarray}
\dot{\delta}_{\rm IDM} + \theta_{\rm IDM} - 3 \dot{\phi} &=& 0 \,,
\\
\dot{\theta}_{\rm IDM} - c_{\rm IDM}^2k^2\delta_{\rm IDM} && \nn\\
+ \mathcal{H}\theta_{\rm IDM} - k^2\psi &=& \Gamma_{\rm IDM-DR} \, \left( \theta_{\rm IDM} - \theta_{\rm DR}\right)\,,\label{eq:idm_eq}
\end{eqnarray}
where $\phi$ and $\psi$ are gravitational potentials, $c_{\rm IDM}$ is the adiabatic IDM sound speed, $\mathcal{H}$ is the conformal Hubble rate,  $k=|\textbf{k}|$ is the comoving wave number of the perturbation, and a dot denotes a derivative with respect to conformal time. On the right-hand side of Eq.~\eqref{eq:idm_eq}, $\Gamma_{\rm IDM-DR}$ is the IDM interaction rate or equivalently the time-derivative of the IDM drag opacity (often written as $\dot{\kappa}$), and $\theta_{\rm DR}$ is the DR velocity divergence.

The equations for the DR density contrast $\delta_{\rm DR}$ and velocity divergence $\theta_{\rm DR}$ are given by
\begin{eqnarray}
\dot{\delta}_{\rm DR} + \frac{4}{3}\theta_{\rm DR} - 4 \dot{\phi} &=& 0 \,,
\\
\dot{\theta}_{\rm DR} - \frac{1}{4}k^2\delta_{\rm DR}&&\nn\\
 {} + k^2\sigma^2_{\rm DR}  - k^2\psi &=& \Gamma_{\rm DR-IDM}\,\left( \theta_{\rm DR} - \theta_{\rm IDM}\right) \,, \label{eq:theta_DR}
\end{eqnarray}
where $\sigma_{\rm DR} = \Pi_{\rm DR,2}/2$ is the shear stress, and $\Gamma_{\rm DR-IDM}$ is related to the drag opacity for DR--IDM coupling.

The two interaction rates $\Gamma_{\rm IDM-DR}$ and  $\Gamma_{\rm DR-IDM}$ can be related using energy-momentum conservation and expanded as power laws in temperature or redshift as follows
 \begin{eqnarray} \label{eq:gamma_dm_dr}
 \Gamma_{\rm IDM-DR} &=& \frac{4}{3}\left( \frac{\rho_{\rm DR}}{\rho_{\rm IDM}}\right)\Gamma_{\rm DR-IDM}  \\
  &=& - \frac{4}{3}(\Omega_{\rm DR}h^2)\,\sum_n a_n  \frac{(1+z)^{n+1}}{(1+z_{\rm D})^n}\,, \nonumber
 \end{eqnarray}
where $h$ is the dimensionless Hubble constant, $\rho_{\rm IDM}$ and $ \rho_{\rm DR}$ are respectively IDM and DR energy densities, and $\Omega_{\rm DR}$ is the DR density in units of the critical density. The coefficients $a_n$ are  constants with units of inverse length, and encapsulate the details of the microscopic model describing the dark sector~\cite{Cyr-Racine:2015ihg, Rubira:2022xhb}. Finally, $z_{\rm D}$ is introduced for normalization and chosen to be $z_{\rm D}=10^7$.

In this work we restrain ourselves to the case $n=0$, which can be realized by assuming a dark sector with interactions governed by an unbroken non-Abelian $SU(N)$ gauge theory~\cite{Buen-Abad:2015ova}, because this scenario can support lower values of $S_8$ \cite{Rubira:2022xhb}. As a convention we take $a_0\equiv a_{\rm dark}$. There are other possible cases $n=2$ and $n=4$ describing, for example, an unbroken $U(1)$ or an interaction with a massive mediator and sterile neutrino as DR, respectively~\cite{Cyr-Racine:2015ihg}. The interaction rate for these two cases has a sharp decrease in time during the radiation dominated era, having an impact only on modes which enter the horizon before the ratio $\Gamma_{\rm IDM-DR}/\mathcal{H}$ falls below unity, leading to a rather sharp suppression of the power spectrum on small scales. It has been shown that these scenarios cannot accommodate lower values of $S_8$~\cite{Archidiacono:2019wdp,Rubira:2022xhb}. On the other hand, for $n=0$ the interaction rate scales in the same way as the expansion rate of the universe during radiation domination, meaning the ratio $\Gamma_{\rm IDM-DR}/\mathcal{H}$ stays constant during this time, impacting many perturbation modes with different scales entering the horizon, and then starts to decrease during matter domination time. This leads to a more gradual suppression of the power spectrum, opening up the possibility to support lower $S_8$ values, see e.g.~Appendix A in~\cite{Rubira:2022xhb}. Another specific property of the non-Abelian dark sector gauge interaction is the presence of a DR self-interaction causing it to behave as a perfect fluid (with suppressed higher multipoles for $l\geq 2$, i.e. $\sigma_{\rm DR} = 0$ in Eq.~\eqref{eq:theta_DR}).

In addition to the cosmological parameters that describe the $\Lambda$CDM model, the IDM--DR model is characterized by  three main parameters (out of which, however, only a subset is relevant here, as detailed below): 
\begin{itemize}
    \item $\xi_{\rm DR}$: the temperature ratio between dark radiation and CMB photons today 
    \be\label{eq:xidef}
\xi_{\rm DR} \equiv \frac{T_{\rm DR}}{T_{\rm CMB}}\biggl|_{z=0}\,,
    \ee
which can equivalently be mapped onto the density parameter of DR as
\be
\omega_{\rm DR} = \Omega_{\rm DR} h^2 = \frac{\eta_{\rm DR}}{2}\zeta\, \xi_{\rm DR}^4\, \Omega_\gamma h^2\,,
\ee
with $\zeta=1$ and $\eta_{\rm DR}=2(N^2-1)$ for a dark sector described by an $SU(N)$ gauge theory, and $\Omega_\gamma$ is the photon density. DR also contributes to the effective number of relativistic species $N_{\rm eff}$, therefore, it can also be parameterized as $\Delta N_{\rm eff}$. We consider the case in which DR behaves as a fluid, as appropriate for an $SU(N)$ dark sector, and therefore set $\Delta N_{\rm eff}= \Delta N_{\rm fluid}$, with
\begin{equation}
\begin{split}
 \Delta N_{\rm fluid} =& \frac{\rho_{\rm DR}}{\rho_{\rm 1\nu}} = \frac{\eta_{\rm DR}}{2} \zeta \xi_{\rm DR}^4  \frac{8}{7}\left( \frac{11}{4}\right)^{4/3}  \\
 &=35.23 \,\xi_{\rm DR}^4\,,
 \end{split}
 \label{eq:deltaN}
\end{equation}
being the ratio between the DR energy density and the energy density of one neutrino family. Throughout this paper, we assume an $SU(3)$ gauge group such that $\zeta=1$ and $\eta_{\rm DR}=16$. These choices have only a minor impact on our conclusions, as they can be absorbed into a rescaling of $\xi_{\rm DR}$.

    \item $f_{\rm IDM}$: the fraction of interacting dark matter
    \be
f_{\rm IDM} \equiv \frac{\Omega_{\rm IDM}}{\Omega_{\rm IDM}+\Omega_{\rm CDM}}\,.
    \ee
    \item $a_{\rm dark}\equiv a_0$: the intensity of the interaction between IDM and DR, as defined in Eq.~\eqref{eq:gamma_dm_dr}.
\end{itemize}

\begin{figure*}
    \begin{center}
        \includegraphics[width=0.335\linewidth]{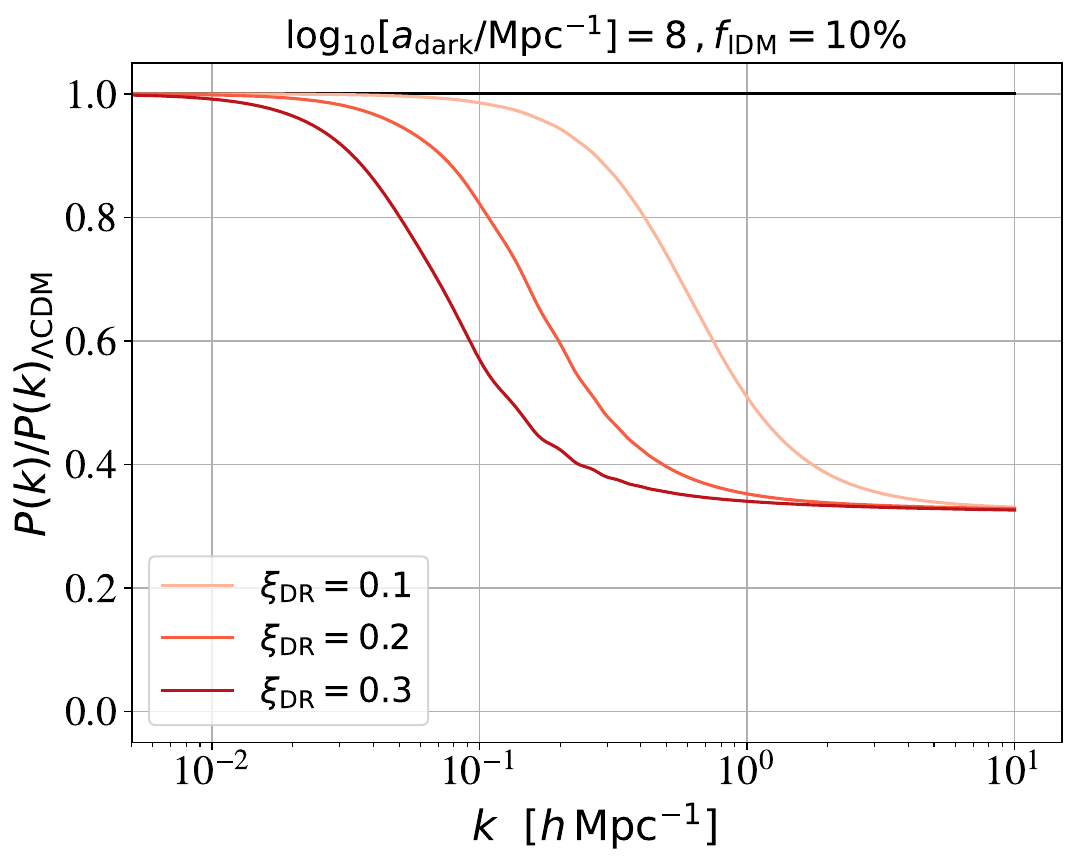}
        \includegraphics[width=0.3\linewidth]{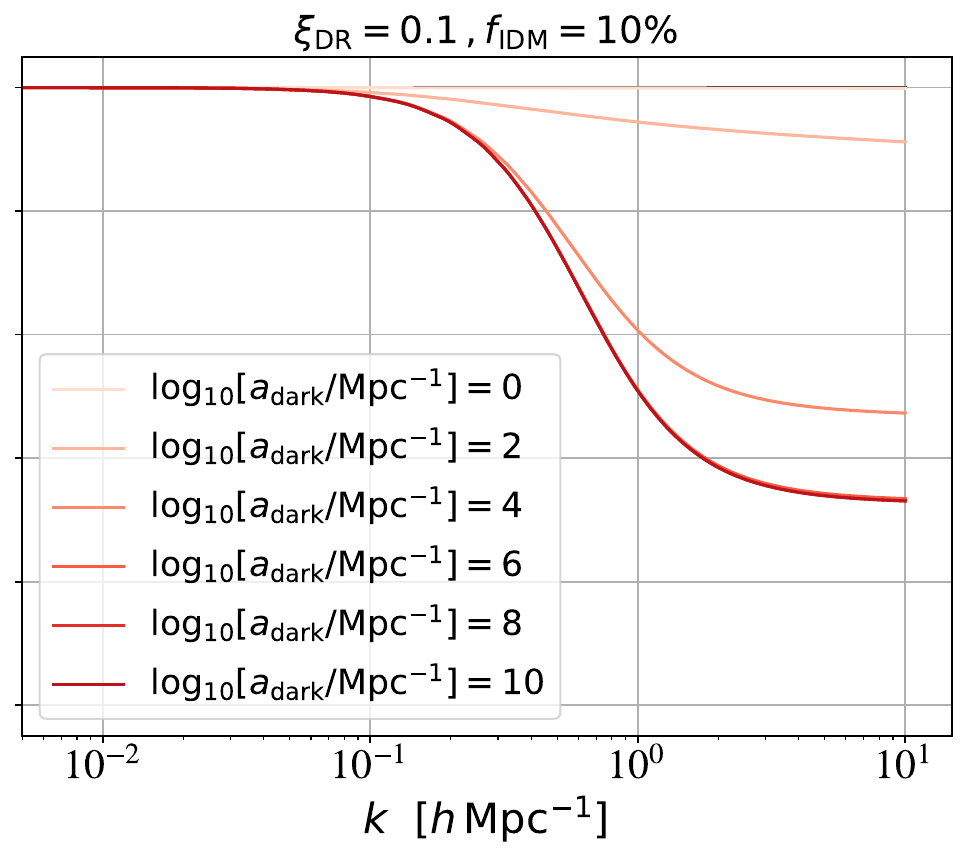}
        \includegraphics[width=0.32\linewidth]{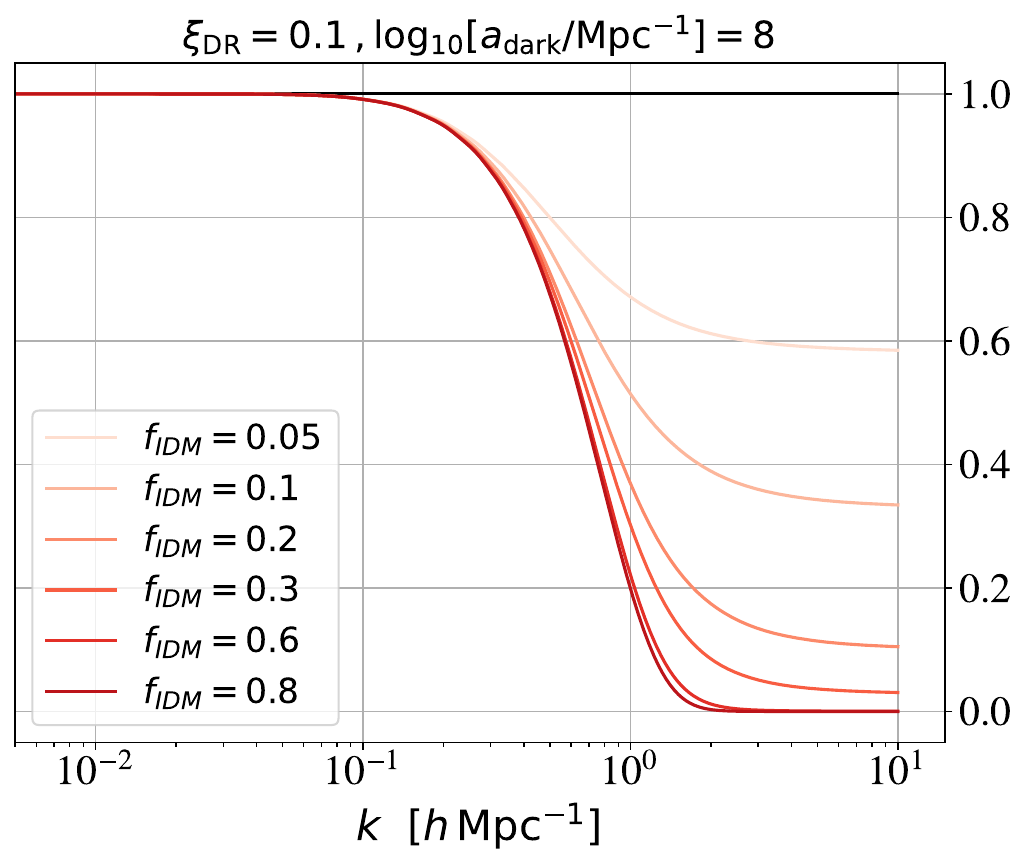}
    \end{center}
    \begin{center}
        \includegraphics[width=0.35\linewidth]{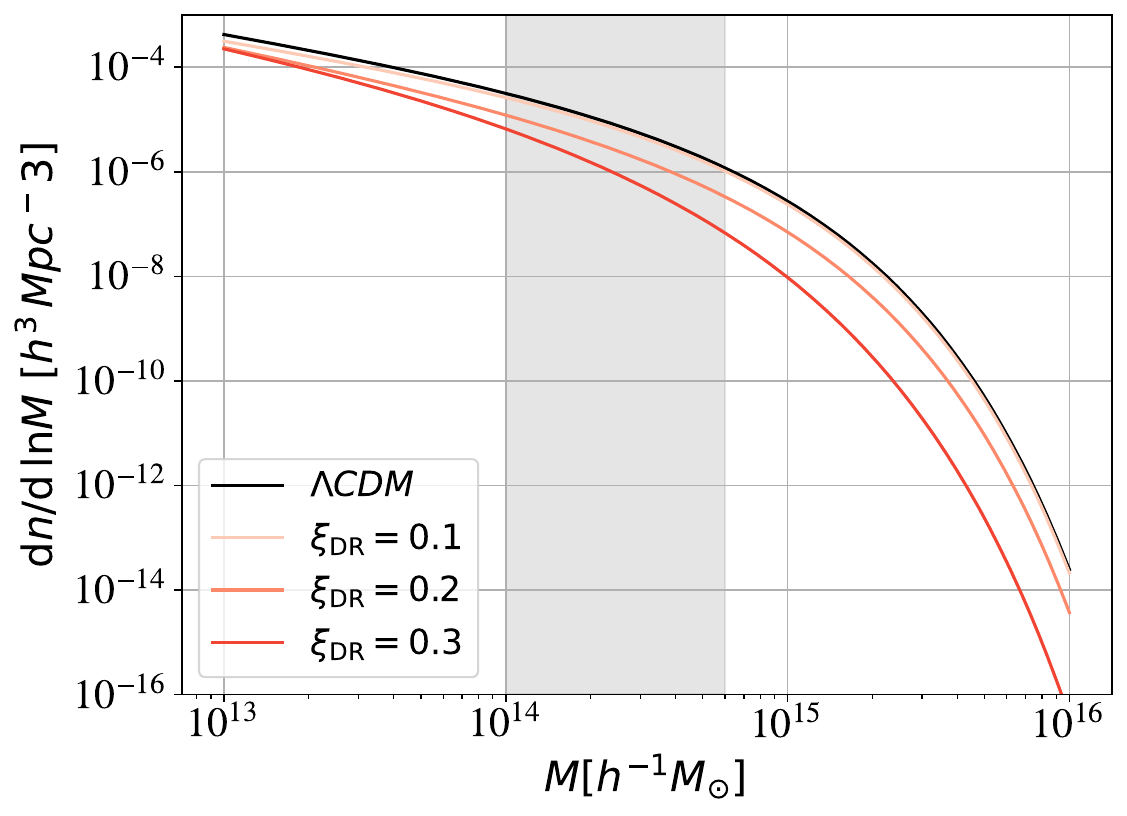}
        \includegraphics[width=0.298\linewidth]{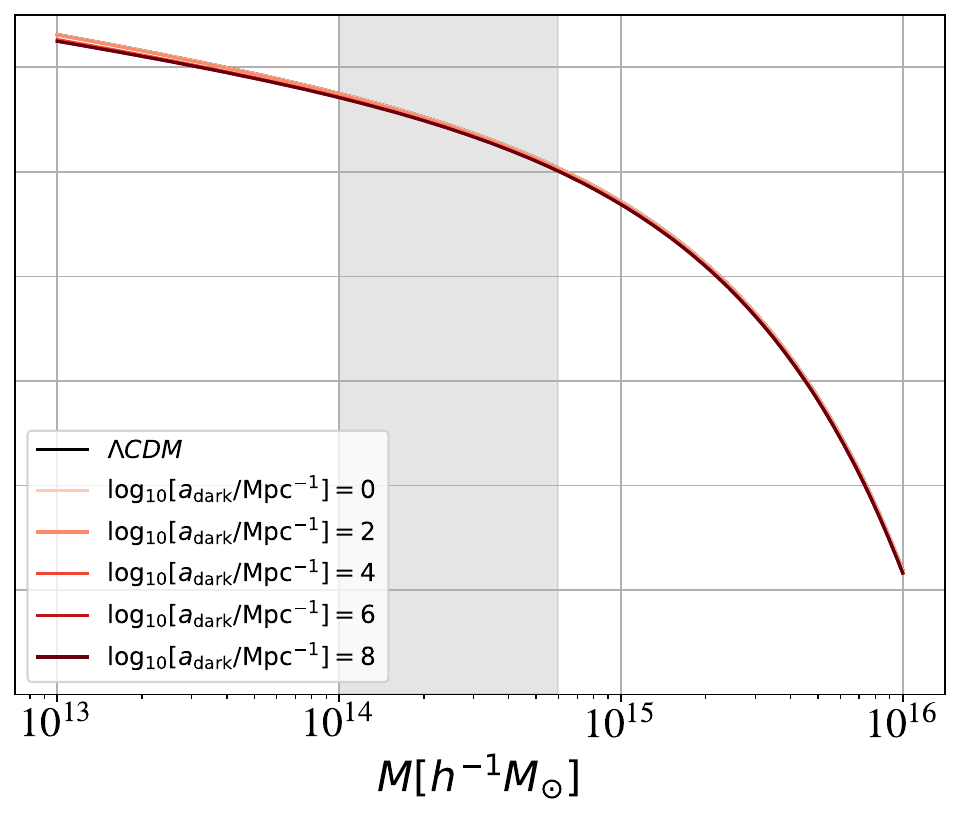}
        \includegraphics[width=0.332\linewidth]{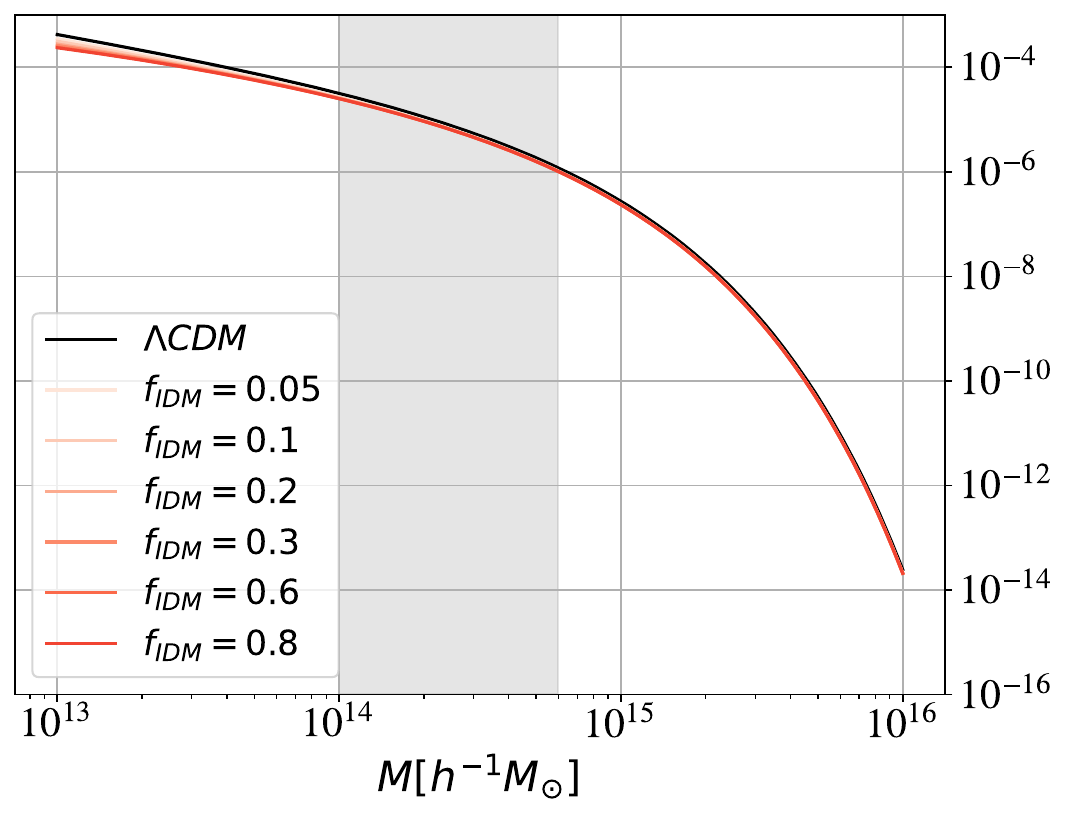}
    \end{center}
\caption{Effect of each IDM--DR parameter on the matter-power spectrum (top) and HMF (bottom). The columns show the impact of varying the DR temperature $\xi_{\rm DR}$ (left), the IDM--DR interaction strength $a_{\rm dark}$ (middle) and IDM fraction $f_{\rm IDM}$ (right), respectively. For the matter power spectrum we show the ratio to the corresponding $\Lambda$CDM spectrum. For the HMF, the gray shaded area refers to the halo masses that  the galaxy cluster samples considered in this work are sensitive to. }
\label{fig:effect_Pk}
\end{figure*}

\subsection{Effect on matter power spectrum and halo mass function}
\label{sec:effect_on_MPS}

In this work we are mostly interested in the effect of IDM--DR on relatively large scales, where precise LSS measurements can probe even small deviations from $\Lambda$CDM. In this context, it is useful to consider two main limits previously discussed in the literature~\cite{Buen-Abad:2017gxg,Rubira:2022xhb}:
\begin{itemize}
    \item {\bf Weak interaction limit}: within this limit, all the DM is interacting with DR ($f_{\rm IDM}=1$), such that to be consistent with CMB and structure formation constraints, the strength of the interaction rate has to be smaller than or comparable to the expansion rate of the universe during radiation domination (RD). 
    \item {\bf Tight coupling limit}: this limit is reached for a sufficiently large interaction rate $\Gamma_{\rm IDM-DR}\gg \mathcal{H}$ during RD (i.e. large $a_{\rm dark}$), and in general viable when only a small fraction of DM is interacting, i.e. for $f_{\rm IDM}\ll 1$. 
\end{itemize}

For concreteness, we mostly focus on the tight coupling limit in the following, but also briefly investigate the impact of lifting this assumption. This limit can be viewed as the dark sector analog of the scenario seen in the visible sector for the interaction between photons and electrons. A mapping between the constraints on the intensity of the interaction parameter $a_{\rm dark}$ to the fine-structure constant $\alpha_{\rm D}$ corresponding to the gauge interaction strength in the dark sector implies that the weak coupling limit corresponds to $\alpha_{\rm D}\sim 10^{-10}-10^{-8}$~\cite{Rubira:2022xhb}. These values are smaller than any coupling known in the Standard Model, while those corresponding to the tight coupling limit can be of comparable size to the electroweak interactions. Additionally, values for $\alpha_{\rm D}$ in the tight coupling limit can also be relevant for DM self-interaction, possibly addressing the small-scale problems \cite{Bullock:2017xww, Tulin:2017ara}.

Due to the interaction with a relativistic component (DR), IDM gains momentum which allows it to escape from overdensities, causing a suppression of structure formation. We demonstrate the impact of the model parameters in Fig.~\ref{fig:effect_Pk}, where we show the dependence of the matter power spectrum $P(k)$ and the HMF within IDM--DR when varying one parameter while fixing the other two. 

The effect of each ETHOS parameter on the matter power spectrum  is shown in the top row of Fig.~\ref{fig:effect_Pk}. The amount of DR (equivalently the temperature ratio $\xi_{\rm DR}$ of DR to the CMB) determines the scale at which the suppression of the matter power spectrum sets in. This is due to two different effects. First, DR has a background effect by contributing to $\Delta N_{\rm fluid}$ and changing the matter-radiation equality time. This effect is small for the values of $\xi_{\rm DR}$ explored here. Second, the main effect comes from the fact that in the tight coupling limit IDM and DR form a dark plasma, and can be treated as a single fluid. During the radiation dominated epoch, $\delta_{\rm IDM}$ tracks $\delta_{\rm DR}$ and both evolve according to a common effective sound velocity term that is primarily sensitive to the value of $\xi_{\rm DR}$ (for more details, see \cite{Buen-Abad:2017gxg}). This situation is analogous to the tightly coupled baryon-photon plasma in $\Lambda$CDM.

The amount of suppression is caused by the other two parameters: the intensity of the interaction $a_{\rm dark}$ and the fraction of IDM $f_{\rm IDM}$. Regarding the intensity of the interaction, we see that for ${\rm log}_{10}[a_{\rm dark}/{\rm Mpc}^{-1}]\gtrsim 6$ the amount of power spectrum suppression becomes independent of the value of $a_{\rm dark}$, indicating that the tight coupling limit has been reached. 
In this tightly coupled limit we are left with $f_{\rm IDM}$ and $\xi_{\rm DR}$ as interesting free parameters (see Appendix~\ref{app:f_idm} for a broader discussion). 
The fraction of IDM is also responsible for the amount of suppression and it can be truncated at $f_{\rm IDM}\sim 20\%$ which suppresses the matter power spectrum by more than $80\%$ and arguably is already excluded by other existing datasets at small scales.  

For the HMF we use the Tinker $N$-body simulation-based model~\cite{Tinker:2008ff}. Akin to previous works for massive neutrinos~\cite{LoVerde:2013lta, Massara:2014kba} and ultralight axion-like particles (ULAs)~\cite{Vogt:2022bwy}, we treat IDM as a biased tracer of CDM, because for small fractions of IDM the halo formation is mainly caused by CDM. The effect of IDM--DR on the HMF is accounted for solely in the matter power spectrum. In the analyses presented here, we follow this approach, leaving a higher fidelity treatment of the impact of IDM--DR on halo formation to a future work. 

The lower row of Fig.~\ref{fig:effect_Pk} shows the impact of each ETHOS parameter on the HMF. The gray shaded area belongs to the halo masses we consider in our samples. Higher values of $\xi_{\rm DR}$ affect the HMF for all halos with $M\gtrsim 10^{13}\,[h^{-1}M_{\odot}]$, which includes the mass range that can be probed with the SPT-3G and CMB-S4 tSZE selected galaxy cluster samples. On the other hand, the effect of $a_{\rm dark}$ and $f_{\rm IDM}$ is relevant only for lowest mass halos $M\lesssim 5\cdot 10^{13}\,[h^{-1}M_{\odot}]$, which are not massive enough to make it into the tSZE selected samples considered here (see Appendix~\ref{app:f_idm}). To distinguish between different values of $a_{\rm dark}$ and $f_{\rm IDM}$ one would need to include a sample of galaxy groups at lower mass or complementary non-cluster constraints on smaller scales.

\section{Galaxy Clusters and Likelihood}
\label{sec:data}

This section contains the description of the galaxy cluster tSZE and weak-lensing datasets employed for this forecast.  We then introduce the so-called observable--mass relations that connect the galaxy cluster observable signatures to the underlying halo masses and then end the section with a presentation of the likelihood.

\subsection{Survey descriptions}

In the following subsections we present the characteristics of the tSZE cluster surveys and the next generation weak lensing (ngWL) surveys.

\subsubsection{Galaxy cluster surveys (SPT-3G and CMB-S4)}
We examine the following two mm-wave surveys that can be used together with optical and NIR followup to produce large tSZE selected cluster samples:
\begin{itemize}
    \item {\bf SPT-3G}
is a mm-wave survey carried out with the South Pole Telescope (SPT) using a third-generation wide-field camera that includes over 15,000 polarization-sensitive bolometers operating at 95, 150 and 220~GHz \cite{SPT-3G:2014dbx}. The survey has been collecting data since 2018 over a sky area of several thousand deg$^2$ and is expected to remain in operation for several years.  The SPT-3G survey is expected to reach noise depths at the few $\mu$K-arcmin level in each band, corresponding to an order of magnitude deeper survey than the SPT-SZ 2500d and a factor of two to three deeper than the second-generation SPT-pol 500d survey~\cite{Austermann:2012ga, Bleem23arXiv231107512B}. We adopt a simplified model of the SPT-3G$\times$ngWL survey with $4,000\,{\rm deg}^2$ of sky coverage split into two regions: a deep main field with $1,500\,{\rm deg}^2$ and a shallower so-called summer field with $2,500\,{\rm deg}^2$. The redshift range of the tSZE selected clusters used in the forecast is $z \in [0.25, 2]$. 
    \item {\bf CMB-S4} 
is the upcoming Stage-4 mm-wave survey that will be carried out using multiple telescopes at the South Pole and in Chile. It is designed to map $\sim 70\%$ of the sky using more than 500,000 polarization-sensitive bolometers to sub $\mu$K-arcmin levels at central CMB frequencies and including frequency coverage ranging from 20 to 270~GHz~\cite{Abazajian:2019eic}. It is expected to yield a tSZE selected cluster sample that is more than one order of magnitude larger than the largest current cluster catalog from ACT~\cite{Abazajian:2019eic,ACT:2020lcv}. To forecast the CMB-S4$\times$ngWL cluster catalog, we follow the survey assumptions described in a previous analysis~\cite{Raghunathan:2021zfi}, which estimated that 75,000 tSZE selected clusters would be identified within the wide survey with a sky coverage of 50\% and reaching a redshift higher than 2.   

\end{itemize} 

We assume here that the cluster samples arise from optical and NIR followup of the tSZE candidate lists employing a technique such as the Multi-Component Matched Filter (MCMF) algorithm \cite{Klein2018MNRAS.474.3324K,Klein2019MNRAS.488..739K}, which has been successfully employed to produce large samples for Planck~\cite{Hernandez-Lang23MNRAS.525...24H} and SPT~\cite{Klein23arXiv230909908K,Bleem23arXiv231107512B} and is currently being used to create an extended ACT cluster sample.  The MCMF algorithm allows one to limit (or reduce) the contamination in a tSZE cluster candidate list in a manner that can be easily accounted for in the cosmological analysis (see discussion below).  This cluster confirmation will benefit from the deep NIR dataset of Euclid as well as the deep optical dataset of Rubin; these are the same ngWL surveys that we envision using for the mass calibration of the tSZE selected cluster samples.

\subsubsection{Next-generation weak lensing (ngWL)}
Ongoing and future surveys like the Euclid mission and the Legacy Survey of Space and Time (LSST) conducted at  the Vera C. Rubin  Observatory will provide a new generation of weak-lensing (ngWL) data. With their large sky coverage and improved instrumentation, they will produce an order of magnitude more WL data than the current available ones with a better quality. For concreteness, we will focus in this work on using ngWL data for the Euclid mission only. The Euclid Wide Survey mission operated by the European Space Agency will cover $15,000\,{\rm deg}^2$ of the extragalactic sky in a time scale of six years. Its imaging and spectroscopy survey in visible and near-infrared bands is designed to improve the estimation of photometric redshifts (photo-$z$s) leading to more precise weak-lensing data \cite{Euclid:2021icp}. It will measure weak-lensing shear profiles of galaxies up to redshift 2 with an average galaxy number density of 30 arcmin$^{-2}$.

\subsection{Observable--mass relations}
\label{sec:observable--mass-relations}
For the two samples presented, the clusters are selected by two observables: the observed {\it detection significance} $\hat\zeta$, which is the detection signal-to-noise ratio (S/N) maximized over all filter scales and cluster sky positions, and the observed {\it richness}, $\hat\lambda$, which quantifies the number of passive galaxies in a cluster. In addition, in the process of optical/NIR confirmation of the ICM selected cluster candidates, a photometric redshift $z$ is measured for each cluster.  Finally, the clusters that overlap the ngWL weak lensing surveys have an associated  projected tangential shear profile $\boldsymbol{g}_\mathrm{t}(R)$. Following the techniques developed within the latest SPT$\times$DES analysis~\cite{Bocquet23arXiv231012213B}, redshift dependent observable--mass scaling relations are used to build a statistical binding between the observables and the underlying halo mass.  

For the samples described below, the clusters are selected to lie above a particular tSZE detection significance threshold $\hat\zeta_\mathrm{min}$ and optical/NIR richness selection threshold $\hat\lambda_\mathrm{min}(z)$, and within a particular redshift range $z_{\rm min}$ and $z_{\rm max}$ as
\begin{eqnarray}
\hat\zeta &>& \hat\zeta_{\rm min}\,,\nonumber \\
\hat\lambda &>& \hat\lambda_{\rm min}(z)\,,\nonumber\\
z_{\rm min} &<& z < z_{\rm max}\,. 
\label{eq:selection}
\end{eqnarray}
There is no selection on the weak lensing observable $\boldsymbol{g}_\mathrm{t}(R)$.

Following the matched filter detection technique employed by SPT~\cite{Melin2006A&A...459..341M}, each cluster is defined to be at the sky location and scale (core radius) that maximize the detection significance, and therefore this significance is somewhat biased.  To correct for this bias, the observed significance $\hat\zeta$ is related to an unbiased or intrinsic tSZE significance $\zeta$ as~\cite{Vanderlinde_2010}
\begin{equation} 
\label{eq:zeta_to_observed}
P(\hat\zeta|\zeta) = \mathcal N\left(\sqrt{\zeta^2 + 3}, 1\right)\,, 
\end{equation}
where ${\cal N}(a,\sigma)$ is a Gaussian distribution with mean $a$ and standard deviation $\sigma$.

The observable-mass relation that connects the mean intrinsic significance $\zeta$ to the halo mass $M_{200c}$ used in describing the HMF is modeled as a power law in mass and in the redshift dependent dimensionless Hubble parameter ($E(z)\equiv H(z)/H_0$) 
\begin{equation}
  \begin{split}
    \langle\ln\zeta\rangle =& \ln\asz + \bsz \ln\left(\frac{M_{200c}}{3\times 10^{14}\,h^{-1}\Msun}\right) \\
    & + \csz \ln\left(\frac{E(z)}{E(0.6)}\right)\,, \label{eq:zetaM}
  \end{split}
\end{equation}
where $\asz$, $\bsz$ and $\csz$ are the scaling relation amplitude, mass trend and redshift trend, respectively. $M_{200c}$ is the halo mass within a region that has an overdensity 200 times larger than the critical density at the cluster redshift.  The pivot redshift and mass for the power law relation are $z=0.6$ and $M_{200c}=3\times10^{14}h^{-1}M_\odot$, respectively.  Observed and simulated clusters exhibit structural variations due to, e.g., the time since last major merger.  This introduces scatter in the observable at any given mass and redshift.  We model the scatter of the intrinsic significance $\zeta$ around the expectation value at a given mass and redshift as log-normal described by its mass and redshift independent RMS (root mean square) variation $\sigma_{\ln\zeta}$.

As previously mentioned, the SPT-3G survey has two fields with different depths that we describe using the field depth parameter $\gamma_{\rm field}$, where 
\be
\zeta_{0,\mathrm{field}} = \gamma_{\rm field} \asz\,. 
\ee
This enables us to combine all clusters within a survey into a single scaling relation with the same mass and redshift behavior, where the differences in field depth simply reflect a rescaling of the normalization parameter \asz. The field depth numbers represent effective inverse noise depths relative to the noise measured in the original SPT-SZ survey \cite{Bleem15ApJS}.

The observed richness $\hat\lambda$ for a particular cluster is the Poisson sampling of an intrinsic richness $\lambda$ as 
        \begin{equation}
        \label{eq:richness_to_observed}
        P(\hat \lambda | \lambda) = \mathcal{N} ( \lambda,  \sqrt{\lambda} )  \, ,
    \end{equation}
where we have used the Gaussian limit of the Poisson distribution, which we consider to be valid for richnesses $\lambda>10$.

The mean intrinsic richness is also modeled as a power law of mass and redshift as
\begin{equation}
  \begin{split}
    \langle\ln\lambda\rangle =& \ln\alambda + \blambda \ln\left(\frac{M_{200c}}{3\times 10^{14}\,h^{-1}\Msun}\right) \\
     &+ \clambda \ln\left(\frac{1+z}{1.6}\right) \,, \label{eq:lambdaM}
  \end{split}
\end{equation}
with the richness-mass scaling relation parameters $\alambda$, $\blambda$ and $\clambda$ corresponding to the amplitude, mass trend and redshift trend.  The log-normal scatter of the richness at a given mass and redshift about the mean richness is modeled using its RMS variation \sigmalnlambda.

These observable--mass scaling relation parameters describe real physical characteristics of the galaxy clusters and are therefore more than so-called nuisance parameters.  
The 8 scaling relation parameters $\ln$\asz, \bsz, \csz, \dsz, $\ln$\alambda, \blambda, \clambda and \dlambda\ 
provide enough freedom to describe the observable--mass relation in existing tSZE selected cluster samples \cite{Bocquet23arXiv231012213B}, and we therefore adopt this same parameter set in forecasting the cosmological constraints for the larger SPT-3G and CMB-S4 samples.  Additional parameters describing further characteristics of the cluster population such as, e.g., mass or redshift dependent intrinsic scatter or correlations among the scatter of different observables can be added as needed when fitting the observed cluster sample.  
The scaling relation parameters are constrained using WL mass calibration as described in Sec.~\ref{sec:WL_mass-calibration}.

The mapping of the weak lensing observable, the reduced tangential shear profile $\boldsymbol{g}_\mathrm{t}(R)$, to the cluster halo mass is handled somewhat differently.  A weak lensing mass $M_\mathrm{WL}$ is extracted from $\boldsymbol{g}_\mathrm{t}(R)$ by fitting a projected NFW model~\cite{Navarro1997ApJ...490..493N} to the one halo portion of the shear profile within the radial range  $500h^{-1}$~kpc~$<R<3.2/(1+z)~h^{-1}$Mpc.  This weak lensing mass is a biased and noisy estimator of the true underlying halo mass $M_{200c}$, and therefore we follow previous publications in creating a statistical binding called the weak lensing to halo mass relation~\cite{Becker2011ApJ...740...25B,Dietrich2019MNRAS.483.2871D,Grandis2021MNRAS.507.5671G},
    \begin{equation}
        \label{eq:WL_mass_rel}
        \begin{split}
         &\left\langle \mathrm{ln} \left( \frac{M_{\mathrm{WL}}}{2 \times  10^{14}\, h^{-1} M_\odot}  \right)  \right\rangle = 
         \aWL \\
        &\hspace{2.5cm}+ 
        \bWL \mathrm{ln} \left( \frac{M_{200c}}{2 \times 10^{14}\, h^{-1} M_\odot} \right) \, ,
        \end{split}
    \end{equation}
where \aWL\ is the logarithmic bias at $M_{200c}=2 \times 10^{14}\, h^{-1} M_\odot$ and \bWL\ is the mass trend of this bias.
The weak lensing mass $M_{\mathrm{WL}}$ exhibits a mass dependent log-normal scatter $\sigma_{\ln\mathrm{WL}}$ about the mean relation given by
    \begin{equation}
    \begin{split}
        \label{eq:WL_mass_var}
      \ln\sigma^2_{\ln\mathrm{WL}} = & \asigmaWL \\
      & + \bsigmaWL  \ln\left( \frac{M_{200c}}{2 \times 10^{14}\, h^{-1} M_\odot} \right) \, ,
    \end{split}
    \end{equation}
where \asigmaWL\ is the logarithm of the variance of $M_\mathrm{WL}$ around $M_{200c}$ at $2 \times 10^{14}\, h^{-1} M_\odot$ and \bsigmaWL\ is the mass trend of this variance.

The parameters of this relation are constrained using a large ensemble of galaxy clusters extracted from numerical structure formation simulations.  Using hydrodynamical simulations~\cite{Nelson:2018uso}, we create mock weak lensing shear profiles $\boldsymbol{g}_\mathrm{t}(R)$ with miscentering and cluster member contamination for and then correct and fit them with NFW profiles in the same manner as the real data are treated, producing $M_{\mathrm{WL}}$ estimates for each halo.  The halo masses $M_{200c}$ are extracted for each cluster from corresponding N-body simulations using the same initial conditions as the hydro simulations.  This is an important detail, because the HMF we use has been extracted from N-body simulations with the same $M_{200c}$ halo mass definition.  These ($M_{\mathrm{WL}},M_{200c})$ pairs are then used to constrain the parameters of the relations presented above.  This whole process is carried out at different redshifts, allowing the bias and scatter of $M_{\mathrm{WL}}$ about the true halo mass $M_{200c}$ to be constrained to some accuracy that depends on the number of simulated clusters studied and the fidelity of the hydrodynamical simulations.  Additional discussion of this approach can be found elsewhere~\cite{Grandis2021MNRAS.507.5671G,Bocquet23arXiv231012213B}.

These remaining parameter uncertainties after this calibration process represent a systematic floor to our ability to constrain the observable-mass relations with weak lensing shear profiles. We return to this discussion in Sec.~\ref{sec:WL_mass-calibration} below.

%
\subsection{Abundance and mass calibration likelihood}
The cluster population is described as independent Poisson realizations of the HMF. The full log-likelihood is given by 
    \begin{equation}
    \begin{split}
      \label{eq:likelihood}
        \ln \mathcal L(\boldsymbol p) = & \sum_i \ln \int_{\hat\lambda_\mathrm{min}}^\infty \dd\hat\lambda\, \frac{\dd^3 N(\boldsymbol p)}{\dd\hat\zeta \dd\hat\lambda \dd z} \Big|_{\hat\zeta_i, z_i} \\
        &- \int_{z_\mathrm{min}}^{z_\mathrm{max}} \dd z \int_{\hat\zeta_\mathrm{min}}^\infty \dd\hat\zeta \int_{\hat\lambda_\mathrm{min}}^\infty \dd\hat\lambda\, \frac{\dd^3 N(\boldsymbol p)}{\dd\hat\zeta \dd\hat\lambda \dd z } \\
        &+ \sum_i \ln\frac{\frac{\dd^4 N(\boldsymbol p)}{\dd\hat\zeta \dd\hat\lambda \dd \boldsymbol g_\mathrm{t} \dd z}
        \Big|_{\hat\zeta_i, \hat\lambda_i, \boldsymbol{g}_{\mathrm{t},i}, z_i}}
        {\int_{\hat\lambda_\mathrm{min}}^\infty \dd\hat\lambda \frac{\dd^3 N(\boldsymbol p)}{\dd\hat\zeta \dd\hat\lambda \dd z}
        \Big|_{\hat\zeta_i, z_i}}
        + \mathrm{const.}\,,
        \end{split}
    \end{equation}
where $\vec p$ refers to model parameters, the sums run over all clusters, and the observables include tSZE significance $\hat\zeta$, richness $\hat\lambda$, redshift $z$ and tangential shear profile $\boldsymbol{g}_{\mathrm{t}}$.  The survey selection is reflected in the integration bounds $z_\mathrm{min}$, $z_\mathrm{max}$, $\hat\zeta_\mathrm{min}$ and $\hat\lambda_\mathrm{min}$.  The first two terms in the likelihood represent the cluster abundance likelihood, which is independent of the weak-lensing data, while the third term represents the information from the mass calibration using the ngWL weak gravitational lensing data.

The differential cluster number $d^3\,N\over \dd\,\mathrm{obs}$ that appears in the first two terms is the differential halo observable function (HOF) in the observable space $\hat\zeta - \hat\lambda - z$ 
    \begin{equation}
    \begin{split}
        \label{eq:HOF_with_3_obs}
        \frac{\dd^3 N (\boldsymbol p)}{\dd\hat\zeta \dd\hat\lambda \dd z } =
        & \int\dd \Omega_\mathrm{s} \iiint   \dd M\, \dd\lambda\, \dd\zeta\, P(\hat\zeta|\zeta) P(\hat\lambda|\lambda) \\
        & P(\zeta, \lambda |M,z,\boldsymbol p) 
         \frac{\dd^2 N (\boldsymbol p,M,z)}{\dd M \dd V} \frac{\dd^2 V (\boldsymbol p,z)}{\dd z \dd \Omega_\mathrm{s}} 
         \, ,
    \end{split}
    \end{equation}
where the integral over the survey solid angle $\Omega_\mathrm{s}$ reduces to the subfield solid angle in a sum over all survey subfields, 
$P(\hat\zeta|\zeta)$ and $P(\hat\lambda|\lambda)$ follow from Eqs.~\eqref{eq:zeta_to_observed} and \eqref{eq:richness_to_observed}, respectively, and $P(\zeta, \lambda |M,z,\boldsymbol p)$ follows from Eqs.~\eqref{eq:zetaM} and \eqref{eq:lambdaM}.  The factors
$\frac{\dd^2 N (\boldsymbol p,M,z)}{\dd M \dd V}= \frac{\dd n(\boldsymbol p,M,z)}{\dd M}$  and $\frac{\dd^2 V (\boldsymbol p,z)}{\dd z \dd \Omega_\mathrm{s}}$ 
are the HMF and
the differential volume element appropriate for the cosmology.

The differential cluster number $\dd\,^4N\over \dd\,\mathrm{obs}$ in the third term of the likelihood is the HOF in the observable space $\hat\zeta - \hat\lambda - \boldsymbol g_\mathrm{t} -  z$
    \begin{equation} 
        \label{eq:HOF_with_4_obs}
        \begin{split}
        \frac{\dd^4 N(\boldsymbol p)}{\dd\hat\zeta \dd\hat\lambda \dd \boldsymbol g_\mathrm{t} \dd z } =
        & \int\dd \Omega_\mathrm{s}\iiiint\dd M\, \dd\zeta\, \dd\lambda\, \dd M_\mathrm{WL}\, \\
        & P(\boldsymbol g_\mathrm{t}|M_\mathrm{WL}, \boldsymbol p)
        P(\hat\zeta|\zeta)
        P(\hat\lambda|\lambda) \\
        & P(\zeta, \lambda, M_\mathrm{WL} |M,z,\boldsymbol p)\\ 
        & \frac{\dd^2 N (\boldsymbol p,M,z)}{\dd M \dd V} \frac{\dd^2 V (\boldsymbol p,z)}{\dd z \dd \Omega_\mathrm{s}} 
        \, .
        \end{split}        
    \end{equation}  
Here the integral over the survey solid angle and the factors $P(\hat\zeta|\zeta)$ and $P(\hat\lambda|\lambda)$ are as described in the previous equation, while $P(\zeta, \lambda, M_\mathrm{WL} |M,z,\boldsymbol p)$ follows from equations Eqs.~\eqref{eq:zetaM}, \eqref{eq:lambdaM} and \eqref{eq:WL_mass_rel}. The factor $P(\boldsymbol g_\mathrm{t}|M_\mathrm{WL}, \boldsymbol p)$ is given by the product of Gaussian probabilities in each radial bin $i$ of the tangential reduced shear profile
    \begin{equation}
    \begin{split}
        \label{eq:lensing_likelihood}
        P(\boldsymbol g_\mathrm{t}|M_\mathrm{WL}, \boldsymbol p) = & \prod_i \left(\sqrt{2\pi}\Delta g_{\mathrm{t},i} \right)^{-1} \\
        & e^{-\frac12 \left(\frac{ g_{\mathrm{t},i} - g_{\mathrm{t},i}(M_\mathrm{WL}, \boldsymbol p)}{\Delta g_{\mathrm{t},i}}\right)^2} \, ,
    \end{split}
    \end{equation}
with the shape noise $\Delta g_{\mathrm{t},i}$.  As previously mentioned, only the radial range $500h^{-1}$~kpc~$<R<3.2/(1+z)~h^{-1}$Mpc, corresponding to the core extracted one halo term region, is used for this fitting.

\section{Mock Data and Analysis}
\label{sec:analysis}

In the following we first discuss the creation of mock cluster catalogs and shear profiles and then describe the analysis of these mock data that leads to our cosmological constraint forecasts. 

\subsection{Generating mock data}
\label{sec:mocks}
To generate the mock catalog for each survey, we compute the matter power spectrum for a given model ($\Lambda$CDM or IDM--DR) with the Boltzmann solver \texttt{CLASS} \cite{Lesgourgues:2011re,Lesgourgues:2011rg}.  The matter power spectrum is then used to calculate the HMF. We use the Tinker~\cite{Tinker:2008ff} simulation-based HMF in the mass range $M \in [10^{13}, 10^{16}]\,h^{-1}\Msun$. 

\begin{table}
    \centering    
    \caption{Cosmological parameters for the two benchmark models examined here. In both models the curvature is set to zero, and the sum of the neutrino masses is $m_{\nu}=0.06\, {\rm eV}$.}
    \label{tab:benchmarks}
    \begin{ruledtabular}
    \renewcommand{\arraystretch}{1.2}
    \begin{tabular}{lcc}
    {\bf Param} & \multicolumn{2}{c}{\bf Benchmark Models} \\
    & $Mock_\mathrm{IDM-DR}$ \quad &   $Mock_{\Lambda\mathrm{CDM}}$\\
    \hline
    $\xi_\mathrm{DR}$ & 0.112 & 0.000 \\
    $\Omega_\mathrm{m}$ & 0.3068 & 0.3166 \\
    $\Omega_\mathrm{b}h^2$ & 0.02550 & 0.02236 \\
    $h$ & 0.6800 & 0.6736 \\
    $\ln(10^{10}A_\mathrm{s})$ & 3.030 & 3.045 \\
    $n_\mathrm{s}$ & 0.9660 & 0.9649 \\
    $\tau_\mathrm{reio}$ & 0.0513 & 0.0544 \\
    \end{tabular}
    \end{ruledtabular}
\end{table}

We generate mock data for two different benchmark cosmologies. The first is based on the IDM--DR model with parameters chosen such that it is compatible with Planck 2018 data, and in addition yields a lower value of $S_8$ due to the interaction of DM with DR,  close to those reported by weak-lensing shear measurements. We choose here to compare to the recent joint analysis of DES-Y3 and KiDS-1000~\cite{Kilo-DegreeSurvey:2023gfr}. The second is based on a $\Lambda$CDM model with input values chosen as the mean parameter posteriors from Planck 2018 temperature and polarization anisotropy without CMB lensing~\cite{Planck:2018vyg}. The cosmological parameters for each benchmark are listed in Table~\ref{tab:benchmarks}. In both cases we consider a zero curvature geometry together with three neutrino species with two massless and one massive neutrino with $m_{\nu}=0.06\, {\rm eV}$. 

To generate the cluster and WL datasets for the two different benchmark cosmologies and the two different survey combinations SPT-3G$\times$ngWL and CMB-S4$\times$ngWL, we begin by using the HMF over the mass range $M \in [10^{13}, 10^{16}]\,h^{-1}\Msun$ and the redshift range of the survey together with the per-field survey solid angle $\Omega_\mathrm{s}$ to estimate the expected number of halos in the surveyed volume as a function of mass and redshift
$\left < N(M,z) \right> =  \frac{\dd^2 N (\boldsymbol p,M,z)}{\dd M \dd V} \dd M \,\Omega_{\rm s} d^2_{\rm p}(z) \dd z$ where $d_{\rm p}(z)$ is the proper distance to redshift $z$ and the HMF contains the comoving abundance of halos (see Eq.~\ref{eq:HOF_with_3_obs}). Taking a Poisson realization of this model, we then cycle through mass and redshift producing the given number of halos within each mass-redshift cell.

Each observed halo of mass $M_{200c}$ and redshift $z$ is then assigned intrinsic observables $\zeta$ and $\lambda$, and thereafter their observable counterparts $\hat\zeta$ and $\hat\lambda$ as described in Sec.~\ref{sec:observable--mass-relations}. 
We generate the observables using the eight $\zeta$- and $\lambda$-mass-$z$ scaling relation parameters fixed to the mean value in the Gaussian priors listed with the mass calibration constraints in Table~\ref{tab:scaling_relations}. To produce the final cluster catalogs for each survey, we apply the survey selection ($\hat\zeta_{\rm min}$, $\hat\lambda_{\rm min}(z)$, $z_{\rm min}$ and $z_\mathrm{max}$) as described in Eq.~\eqref{eq:selection}. For the redshift dependent richness threshold we adopt the values obtained for the recent SPT$\times$DES $\Lambda$CDM analysis~\cite{Bocquet23arXiv231012213B}.

For SPT-3G, the full region is covered by the ngWL surveys Euclid and Rubin ($\Omega_\mathrm{s}=4,000 \,{\rm deg}^2$) with a redshift range of $0.25<z\lesssim 2$. Following the recent SPT$\times$DES analysis~\cite{Bocquet23arXiv231012213B}, we use two selection thresholds for the significance, one in each of the two fields:   $\hat\zeta_\mathrm{min}=5$ for the summer field and $\hat\zeta_\mathrm{min}=4.5$ for the main field. The main field is described by a field depth parameter $\gamma_{\rm field}=3.5$, and the shallower summer field has $\gamma_{\rm field}=1.5$. 
With these survey specifications we obtain around 4,000 clusters for ${\it Mock}_{\,\rm IDM-DR}$ and 6,400 clusters for ${\it Mock}_{\,\rm \Lambda CDM}$.

\begin{table}
    \centering    
    \caption{Scaling relation parameters with priors and constraints.  
    Priors on the weak lensing to halo mass relation are taken from the previous SPT$\times$DES analysis~\cite{Bocquet23arXiv231012213B} with uncertainties reduced by a factor of two.
    Constraints on the tSZE significance and richness observable--mass scaling relation parameters are derived from the WL mass calibration. ${\cal N}(a,\sigma)$ is a Gaussian distribution with mean $a$ and standard deviation $\sigma$.  }
    \label{tab:scaling_relations}
    \begin{ruledtabular}
    \renewcommand{\arraystretch}{1.2}
    \begin{tabular}{lcc}
    {\bf $M_{\rm WL}-M_{200c}$} & \multicolumn{2}{c}{Priors} \\
    \hline
    $\aWL$ & \multicolumn{2}{c}{$\mathcal{N}(-0.05,\, 0.01)$} \\
    $\bWL$ & \multicolumn{2}{c}{$\mathcal{N}(1.029,\, 0.009)$}  \\
    $\asigmaWL$ & \multicolumn{2}{c}{$\mathcal{N}(\ln(0.212)^2,\, (0.110)^2)$}   \\
    $\bsigmaWL$ & \multicolumn{2}{c}{$\mathcal{N}(-0.226,\, 0.300)$}\\
    \hline
    {\bf $\zeta-M_{200c}-z$} &
    \multicolumn{2}{c}{Constraints} \\
    \hline
     & CMB-S4$\times$ngWL \quad &   SPT-3G$\times$ngWL\\
    $\ln$\asz & $\mathcal{N}(0.960,\, 0.025)$ & $\mathcal{N}(0.96,\, 0.031)$\\
    \bsz & $\mathcal{N}(1.500,\, 0.038)$  & $\mathcal{N}(1.50,\, 0.041)$ \\
    \csz & $\mathcal{N}(0.500,\, 0.099)$  & $\mathcal{N}(0.50,\, 0.17)$ \\
    \dsz & $\mathcal{N}(0.200,\, 0.017)$  & $\mathcal{N}(0.200,\, 0.026)$ \\
    \hline
    {\bf $\lambda-M_{200c}-z$} & \multicolumn{2}{c}{Constraints} \\ 
    \hline
         & CMB-S4$\times$ngWL \quad &   SPT-3G$\times$ngWL\\
    $\ln$\alambda & $\mathcal{N}(4.250,\, 0.016)$ & $\mathcal{N}(4.25,\, 0.02)$\\
    \blambda & $\mathcal{N}(1.000,\, 0.029)$ & $\mathcal{N}(1.000,\, 0.031)$  \\
    \clambda & $\mathcal{N}(0.000,\, 0.070)$ & $\mathcal{N}(0.00,\, 0.12)$ \\
    \dlambda & $\mathcal{N}(0.200,\, 0.009)$ & $\mathcal{N}(0.20,\, 0.02)$ 
    \end{tabular}
    \end{ruledtabular}
\end{table}

The overlapping region for CMB-S4 and the ngWL survey Euclid is roughly $\Omega_\mathrm{s}=10,100\,{\rm deg}^2$ and even larger for Rubin. We adopt the solid angle of the overlap with Euclid and a  redshift range of $0.1<z\lesssim 2$ (i.e., $z_\mathrm{min}=0.1$).  The gains from adopting the Rubin overlap are marginal so we do not explore them here. Although CMB-S4 reaches tSZE sensitivities where we expect to find clusters at redshifts higher than 2, we restrict our analysis to clusters with $z\lesssim 2$. Cluster cosmology using clusters at $z>2$ has not yet been attempted, and so we focus here on the better understood regime $z<2$. The significance threshold for CMB-S4 is taken to be $\hat\zeta_\mathrm{min}=5$ and the field depth parameter is $\gamma_{\rm field}=4.0$.
The survey defined in this way includes approximately 24,000 clusters for ${\it Mock}_{\,\rm IDM-DR}$ and 32,000 clusters for ${\it Mock}_{\,\rm \Lambda CDM}$. 

In addition to the tSZE and optical/NIR observables, each selected cluster halo is also assigned a mass estimate $M_\mathrm{WL}$ following Eqs.~\eqref{eq:WL_mass_rel} and ~\eqref{eq:WL_mass_var}.  Using $M_\mathrm{WL}$, each mock cluster is then provided with a mock shear profile consistent with the NFW model that is used in fitting the shear profiles during the analysis. The lensing source galaxies are distributed within ten tomographic redshift bins extending over the redshift range 0 to 2.2. Profile measurement uncertainties are assigned using shape noise assuming intrinsic shear variance of $(0.3^2)$ and a lensing source galaxy surface density of $30~\mathrm{arcmin}^{-2}$. The amplitude of the shear profiles for each tomographic bin is calculated using the appropriate $\Sigma_\mathrm{crit}$ factor, where the weighting factor is set to zero for those bins at redshifts less than the cluster redshift.
Cluster member contamination of the source galaxy sample is added, consistent with measurements extracted for tSZE selected clusters within the SPT$\times$DES shear dataset.

The priors on the weak lensing to halo mass relation are listed in Table~\ref{tab:scaling_relations} and discussed further in Sec.~\ref{sec:WL_mass-calibration} below.  In producing the mock data the central values of each parameter are adopted, and this relation is applied to mock clusters at all redshifts.

\subsection{Analysis}

The Markov chain Monte Carlo (MCMC) analysis is carried out using {\bf CosmoSIS}\footnote{\url{https://cosmosis.readthedocs.io/}} \cite{Zuntz:2014csq} with the Polychord sampler \cite{Handley:2015fda,Handley:2015vkr}. We cross-check that the MultiNest sampler \cite{Feroz:2008xx,Feroz:2007kg,Feroz:2013hea} yields consistent results for all posterior distributions. 

In recent weak lensing informed cluster abundance analyses~\cite{SPT:2018njh, Bocquet23arXiv231012213B}, the full log-likelihood containing both the cluster abundance and the weak lensing mass calibration have been iterated simultaneously. This is optimal, because the weak-lensing mass calibration is dependent on the distance--redshift relation, which is sensitive to the cosmological parameter, e.g., $\Omega_\mathrm{m}$. However, for the forecast analysis performed here where we include both cluster datasets and Planck 2018 anisotropy data, the two pieces of the likelihood can be separated, leading to gains in computing time. In a related analysis~\cite{Vogt23arXiv231012213B}, it has been shown that the dependence of the mass calibration on cosmology for the cluster samples considered here is negligible, underscoring the validity of separating the analysis.
Therefore, we perform the  mass calibration independently, as described in the following Sec.~\ref{sec:WL_mass-calibration}, and adopt the resulting parameter posteriors as priors for a stand-alone cluster abundance likelihood analysis as described in Sec.~\ref{sec:abundance-analysis}. 

\subsubsection{WL mass calibration analysis}
\label{sec:WL_mass-calibration}

To carry out the WL mass calibration, 
the mock shear profiles are analyzed in the same manner as the observed DES based shear profiles employed in the recent SPT$\times$DES analysis \cite{Bocquet23arXiv231012213B}. They are corrected for cluster member galaxy contamination and 
fit using a projected NFW
density profile over a region that excludes the two-halo term dominated cluster infall region and the central 500$h^{-1}$kpc where the cluster profile is particularly sensitive to feedback effects, hydrodynamical modeling and the choice of cluster center. This fit results in a WL mass estimate \Mwl\ for each cluster.

 As discussed previously in Sec~\ref{sec:mocks}, the \Mwl\ measurement is a biased and noisy estimate of the underlying halo mass $M_{200c}$.
 We account for this source of systematic and stochastic uncertainty by marginalizing over the uncertainties in the parameters of the weak lensing to halo mass relation described in 
Eqs.~\eqref{eq:WL_mass_rel} and~\eqref{eq:WL_mass_var}).  
 
 This approach then accounts for mismatch between complex cluster morphologies and the simple NFW shear profile used to estimate \Mwl\ as well as for the impact of uncertainties in cluster mis-centering and cluster member galaxy contamination of the shear source galaxy sample.  Moreover, systematic uncertainties on the photometric redshifts of the source sample and on the shear measurements themselves can be included.  Finally, there are additional systematics due to the uncertainties in the subgrid physics used in the hydrodynamic simulations that have been employed in constraining this $M_{\rm WL}-M_{\rm halo}$ relation.

 For the WL calibration used here, we adopt the systematic requirements for the source galaxy photometric redshifts and shear measurements from the ngWL surveys. These requirements are so tight in comparison to existing shear dataset from, e.g., DES, that those systematics become subdominant in our mass calibration analysis. For the remaining systematic uncertainties associated with model mismatch, mis-centering, member galaxy contamination and hydrodynamical subgrid physics, we find that hydrodynamical subgrid physics uncertainties of $\sim$2\%~\cite{Grandis2021MNRAS.507.5671G} dominate. Because it is not clear how much improvement to expect in the remaining systematics between now and the time when the actual samples are analyzed, we adopt a conservative approach where the systematics are half as large ($\sim1$\%) and an optimistic approach where the systematics are 10 times smaller ($\sim0.2$\%). As discussed in the companion paper~\cite{Vogt23arXiv231012213B}, the ngWL datasets are essentially shape noise dominated in both the conservative and optimistic scenarios, making the exact level of systematic uncertainties we include unimportant for our forecasts. Therefore, we choose to adopt the conservative approach with an uncertainties on \aWL\ of 1\%. The mass trend in the logarithmic bias and the scatter parameters of the weak lensing to halo mass relation are taken from the recent SPT$\times$DES analysis with uncertainties reduced by a factor of two~\cite{Bocquet23arXiv231012213B}. All parameter priors are listed in Table~\ref{tab:scaling_relations}.

Because the cluster by cluster WL mass calibration analysis would have taken too long for the samples analyzed here, we adopt an approach of using only 1000 randomly selected clusters for the mass calibration in each sample, where the shape noise is reduced to maintain the information content expected from an analysis of each full sample (further discussion appears elsewhere~\cite{Vogt23arXiv231012213B}). The analysis leads to tight posteriors on the parameters of the  observable--mass relations 
as shown in Table~\ref{tab:scaling_relations} for SPT-3G$\times$ngWL and CMB-S4$\times$ngWL. We adopt these parameter constraints as Gaussian priors in the analysis of the cluster abundance described in the following section. Note that to account for possible halo mass shape changes due to IDM and their impact on the WL mass calibration, separate structure formation simulations could be employed.  Ror the forecasts presented here we simply assume that the WL systematic uncertainties are the same for both IDM and CDM.

\subsubsection{Cosmology analysis}
\label{sec:abundance-analysis}

Our analysis here is for the cluster counts from SPT-3G and CMB-S4 with mass information from future ngWL surveys like Euclid and Rubin. We use  the likelihood for cluster abundance as described above, which is based upon the most recent SPT$\times$DES analysis \cite{Bocquet23arXiv231012213B} and the likelihood for CMB Planck 2018 TT,TE,EE data available in CosmoSIS. An analysis using Planck data for this model has already been performed~\cite{Archidiacono:2019wdp,Rubira:2022xhb}, but we include it here for comparison. Therefore, we carry out a cosmology analysis for
\begin{itemize} 
    \item Planck 2018 (TT,TE,EE)
    \item SPT-3G (tSZE clusters) $\times$ ngWL (WL masses) 
    \item CMB-S4 (tSZE clusters) $\times$ ngWL (WL masses)
\end{itemize}
In the following we refer to these datasets as Planck, SPT-3G$\times$ngWL, and CMB-S4$\times$ngWL, respectively. In addition, we consider also the combination of Planck with either SPT-3G$\times$ngWL or CMB-S4$\times$ngWL to assess the additional constraining power of future cluster data.
 We explore the parameters of $\Lambda$CDM adding the parameter $\xi_{\rm DR}$ for the IDM--DR model. We use flat priors on these as shown in Table~\ref{tab:params}. Priors on observable--mass scaling relation parameters are summarized in Table~\ref{tab:scaling_relations}. Cluster counts are not sensitive to $\Omega_{\rm b} h^2$ and provide only weak constraints on $h$. Therefore, when using cluster data alone, we adopt Gaussian priors on them, as given in Table~\ref{tab:params}.

The prior on $\xi_{\rm DR}$ comes from the constraint on extra (non-interacting) radiation $\Delta N_{\rm eff}$ from Planck, that can be viewed as an upper limit in the interacting case. As previously stated, we work in the tight coupling limit, meaning we fix the value of the interaction intensity 
to be large enough such that our results do not depend on its precise value. Specifically, we use ${\rm log}_{10}[a_{\rm dark}/{\rm Mpc^{-1}}]=8$. For simplicity and to avoid parameter projection effects, we also fix the IDM fraction to be $f_{\rm IDM}=10\%$. As discussed in Sec.~\ref{sec:effect_on_MPS}, the matter power spectrum and HMF are rather insensitive to its precise value within the relevant regime for the cluster samples considered here. Nevertheless, an analysis with $a_{\rm dark}$ and $f_{\rm IDM}$ included as free parameters is shown in Appendix~\ref{app:f_idm}.

\begin{table}
\caption{\label{tab:params}%
Summary of parameter priors adopted in the analysis. The $\mathcal{U}$ stands for a uniform distribution. $\tau_{\rm reio}$ is considered only when adding CMB data, and Gaussian priors on $\Omega_{\rm b} h^2$ and $h$ are added when using cluster data alone, with mean $c$ standing for the value of $\Omega_{\rm b} h^2$ and $d$ for the value of $h$ as specified for either ${\it Mock}_{\,\rm IDM-DR}$ or ${\it Mock}_{\,\rm \Lambda CDM}$, respectively.
}
\begin{ruledtabular}
\renewcommand{\arraystretch}{1.2}
\begin{tabular}{lc}
\textrm{Parameter}&
\textrm{Prior}
\\
\colrule
$\xi_{\rm DR}$ & $\mathcal{U}(0.001,\, 0.5)$ \\
$\Omega_{\rm m}$ & $\mathcal{U}(0.2,\, 0.4)$ \\
$\Omega_{\rm b} h^2$ & $\mathcal{U}(0.018, \,0.027)\big(\times \mathcal{N}(c,\,0.00015)\big)$\\
$h$ & $\mathcal{U}(0.6,\, 0.8)\big(\times \mathcal{N}(d,\,0.006)\big)$\\
${\rm ln}\,(10^{10}A_{\rm s})$ & $\mathcal{U}(1.0,\, 4.0)$\\
$n_{\rm s}$ & $\mathcal{U}(0.94,\, 1.00)$\\
$\tau_{\rm reio}$ & $\mathcal{U}(0.04,\, 0.08)$\\
\end{tabular}
\end{ruledtabular}
\end{table}

\section{Results and discussion}
\label{sec:results}
We present here the results for the forecast sensitivity to an interacting dark sector of either the SPT-3G or the CMB-S4 tSZE galaxy cluster samples combined with weak-lensing mass calibration from a Euclid- or Rubin-like ngWL survey. 
We present results of our MCMC analysis within the IDM--DR and CDM context for the two benchmark models (specified in Table~\ref{tab:benchmarks}) in Sec.~\ref{IDM} and Sec.~\ref{LCDM}, respectively. In addition, in Sec.~\ref{sec:neutrinos} we discuss the interplay with massive neutrinos. While our MCMC runs include all parameters mentioned previously, we display only the results for $\xi_{\rm DR}$, $\Omega_{\rm m}$, and $S_8$. The full results are displayed in Appendix~\ref{app:cosmo24_all}.

Cosmological probes can constrain the parameter combination $\sigma_8 (\Omega_m/0.3)^{\alpha}$ for different values of $\alpha$ depending on the dataset and the specifics of the survey. For the surveys in this work, we find that CMB-S4 is better at constraining the parameter combination for $\alpha=0.18$, and SPT-3G for $\alpha=0.2$. Nevertheless, we will display results in terms of the parameter combination $S_8=\sigma_8 (\Omega_m/0.3)^{0.5}$ to ease comparison with previously published shear analyses and to address the corresponding tension.
\subsection{IDM--DR benchmark}
\label{IDM}
\begin{figure*}
    \begin{center}
        \includegraphics[width=0.49\linewidth]{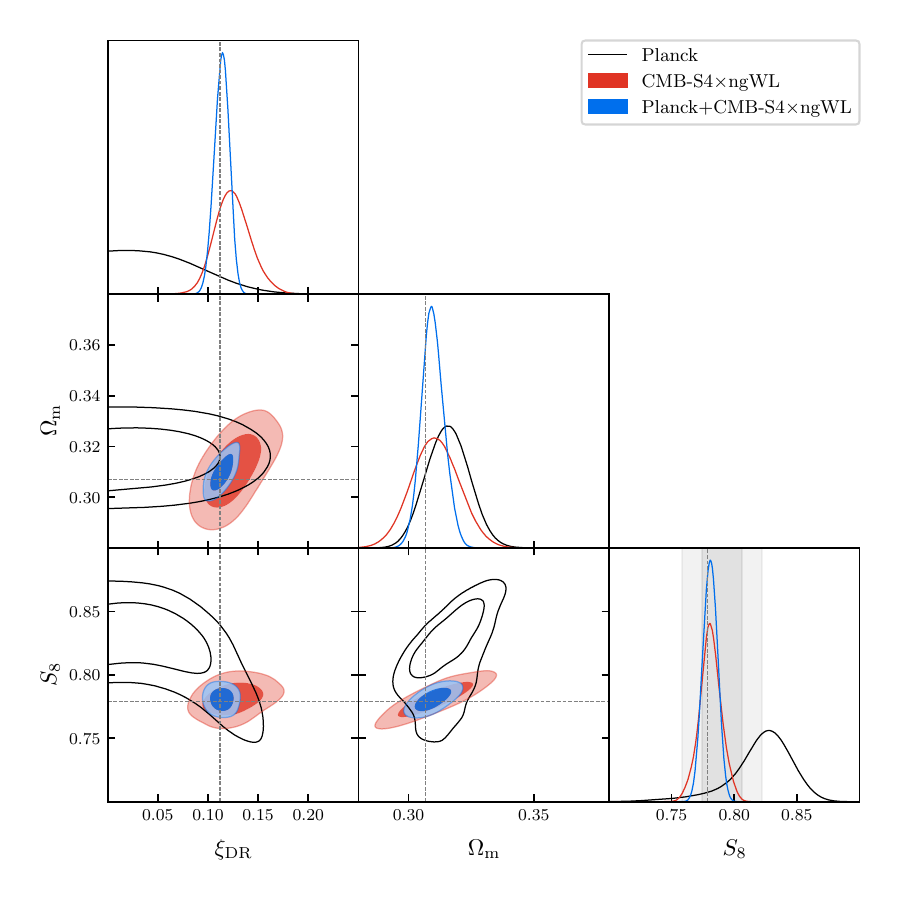}
        \includegraphics[width=0.49\linewidth]{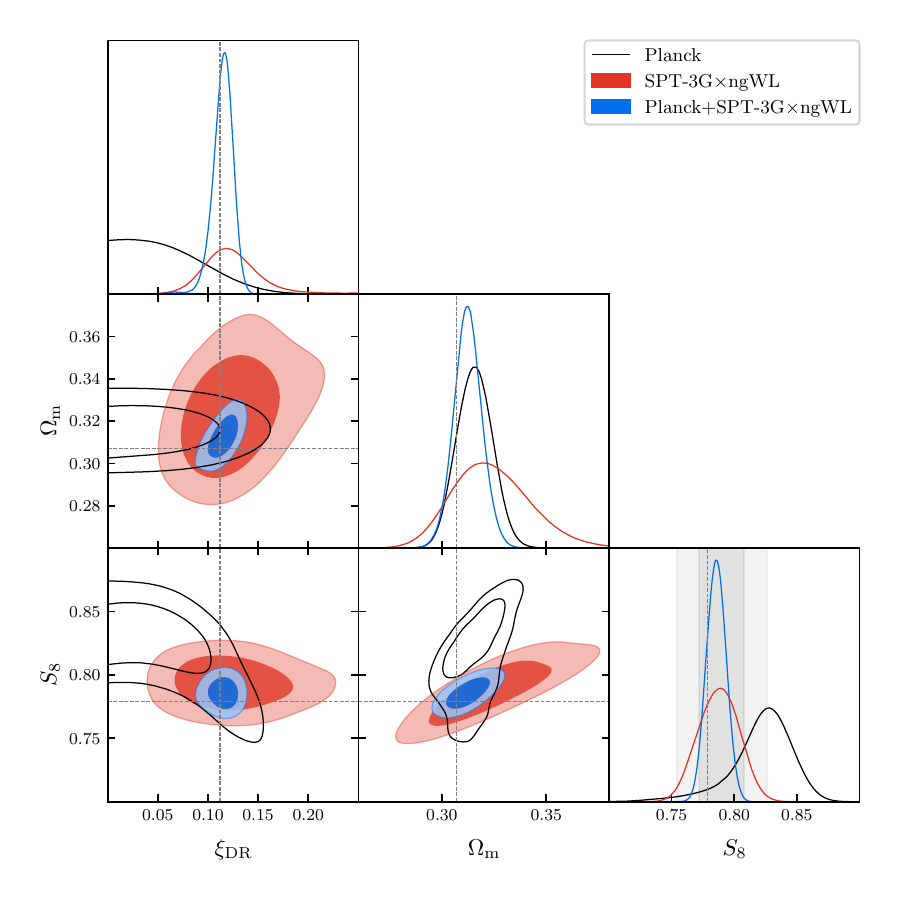}
    \end{center}   
\caption{Posteriors within the IDM--DR model when assuming the ${\it Mock}_{\,\rm IDM-DR}$ benchmark model. On the left, we show the forecast for CMB-S4$\times$ngWL and on the right SPT-3G$\times$ngWL. Red contours show the forecast from cluster abundance only, while blue corresponds to a combination with Planck 2018. For comparison, the black lines show current constraints from Planck 2018 only. The gray shaded area refers to the results from the joint analysis of DES-Y3 and KiDS-1000~\cite{Kilo-DegreeSurvey:2023gfr}, and the dotted lines refer to the input values of the benchmark model used to generate the mock data.}
\label{fig:IDM_cosmo24}
\end{figure*}

We first discuss the IDM--DR benchmark model chosen to be compatible with Planck 2018 data and additionally yielding a low value of $S_8$ due to dark sector interactions. This scenario is relevant if the hints for low values of $S_8$ from weak lensing shear and current cluster measurements turn out to be true. The results of the analysis using ${\it Mock}_{\,\rm IDM-DR}$ data are shown in Fig.~\ref{fig:IDM_cosmo24} for the dataset combinations described in Sec~\ref{sec:analysis}. The values of the parameters at 68\%  or upper limits at 95\% credibility level (CL) with their relative error are summarized in Table~\ref{tab:results}. 
We highlight that all input values are recovered within $1\sigma$ for all the results presented in this work, indicating self-consistency in the theoretical description, the mock generation and the likelihood analysis.

In particular, the temperature of the dark sector parameterized by $\xi_{\rm DR}$ is recovered and tightly constrained from galaxy cluster catalogs with a relative error of 14\% and 42\% for CMB-S4$\times$ngWL and SPT-3G$\times$ngWL, respectively. This means the cluster abundance is able to distinguish the IDM--DR benchmark model from $\Lambda$CDM, which corresponds to the limit $\xi_{\rm DR}\to 0$. Thus, both SPT-3G and CMB-S4 cluster counts will be able to test whether dark sector interactions are an explanation of the $S_8$ tension, when combined with ngWL mass measurements.

The power of CMB-S4$\times$ngWL 
in constraining the value of $\xi_{\rm DR}$ is much higher than that of SPT-3G$\times$ngWL. This could be traced to the difference in the size of catalogs, where for this model CMB-S4 will be able to identify around 24,000 clusters compared to 4,000 clusters for SPT-3G. Additionally, the wide overlapping region of the CMB-S4 and ngWL surveys leads to better mass calibration, which means less uncertainty in the observable--mass relations (see Table~\ref{tab:scaling_relations}).
Both cluster samples yield a comparable constraining power on $S_8$ at a precision of about $1\%$ for CMB-S4$\times$ngWL and $2\%$ for SPT-3G$\times$ngWL (see Table~\ref{tab:results}). As is clear in Fig.~\ref{fig:IDM_cosmo24}, combining cluster catalogs with Planck temperature and polarization anisotropy likelihoods strengthens the constraining power on the parameters and dramatically so in the case of the smaller SPT-3G cluster sample. 

Notably, current Planck 2018 data are not able to discriminate the IDM--DR benchmark model from $\Lambda$CDM (see black contour lines in Fig.~\ref{fig:IDM_cosmo24}). Instead, Planck only yields an upper limit $\xi_{\rm DR} \leq 0.13$ (see Table~\ref{tab:results}). The same is true for combinations of Planck with BAO or BOSS galaxy clustering data~\cite{Archidiacono:2019wdp,Rubira:2022xhb}. Thus, future cluster abundance measurements will significantly enhance the sensitivity to interacting dark sectors. We find that CMB-S4 cluster catalogs together with ngWL mass calibration will be able to decisively test IDM--DR as an explanation of the $S_8$ tension.

\subsection{$\Lambda$CDM benchmark}
\label{LCDM}

\begin{figure*}
    \begin{center}
        \includegraphics[width=0.49\linewidth]{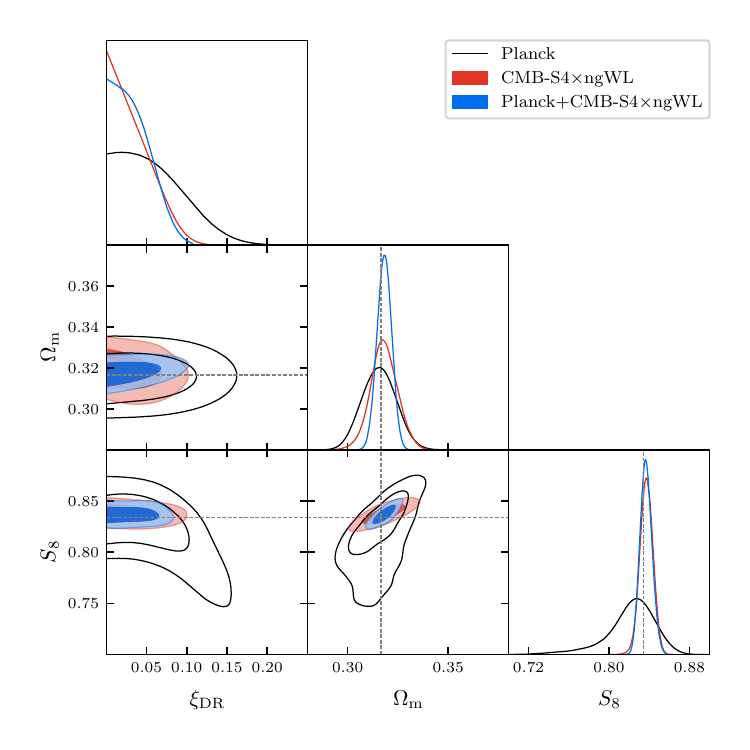}
        \includegraphics[width=0.49\linewidth]{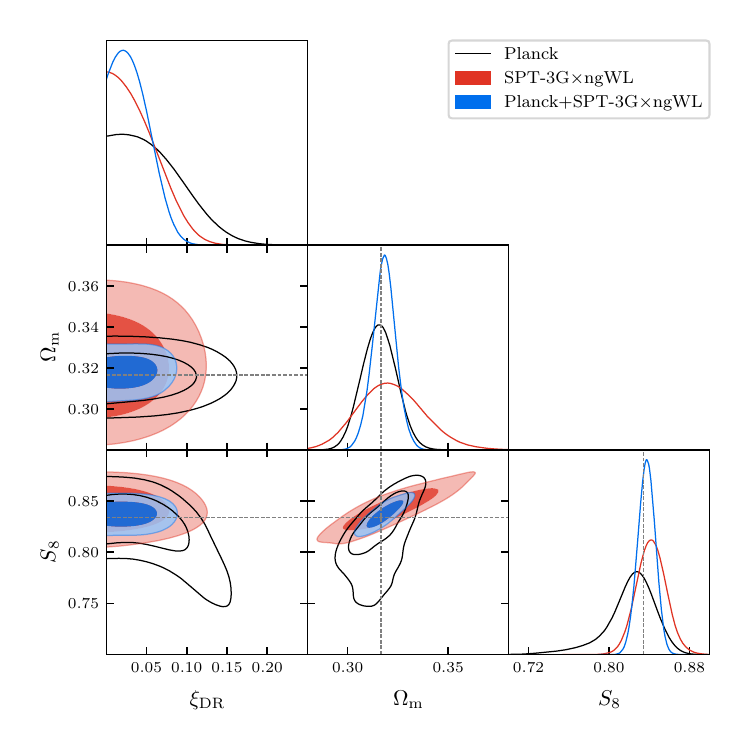}
    \end{center}
\caption{Same as Fig.~\ref{fig:IDM_cosmo24} for the ${\it Mock}_{\,\rm \Lambda CDM}$ benchmark model, corresponding to a $\Lambda$CDM model with Planck 2018 parameters.}
\label{fig:LCDM_tcl}
\end{figure*}

Let us now consider the scenario that the true cosmological model is described by $\Lambda$CDM with parameters compatible with Planck 2018 and a correspondingly high value of $S_8$, which is relevant if the current hints for low $S_8$ from weak lensing shear and other measurements are not confirmed in future analyses. In this case one may ask how future cluster abundance data will enable improved constraints on the IDM--DR model.
The results of the MCMC sampled posteriors for ${\it Mock}_{\,\rm \Lambda CDM}$ are shown in Fig.~\ref{fig:LCDM_tcl} and the values of the parameters at 68\%  or upper limits at 95\% CL are given in Table~\ref{tab:results}.

If the true cosmology is $\Lambda$CDM we find that cluster abundances will allow us to reach an upper bound on the value of $\xi_{\rm DR}$ of 0.08 and 0.09 for CMB-S4$\times$ngWL and SPT-3G$\times$ngWL, respectively, at 95\% CL. The degeneracy between $S_8$ and $\xi_{\rm DR}$ that is present in Planck 2018 CMB data is lifted, and the possibility of a lower value of $S_8$ due to dark sector interactions can be excluded within this scenario. 
 
 This means if $\Lambda$CDM accurately describes our Universe, future cluster counts will improve the constraints on $\xi_{\rm DR}$ over Planck 2018 combined with BOSS, BAO and galaxy clustering by about a factor $1.5-2$. Note that this implies an improvement in the upper limit on the abundance of DR $\Delta N_{\rm fluid}\propto \xi_{\rm DR}^4$ by a factor $5-10$. Using Eq.~\eqref{eq:deltaN}, the upper bound on $\xi_{\rm DR}$ translates 
 to an upper bound on $\Delta N_{\rm fluid}$ of 0.0014 and 0.0023  
 for CMB-S4$\times$ngWL and SPT-3G$\times$ngWL, respectively, for $SU(3)$ theory. 
 
 Moreover, future cluster abundance data will allow us to reduce the uncertainty on $S_8$ by about a factor of two compared to CMB data from Planck (with a relative error of $0.7\%-1.7\%$ for clusters and $2.4\%$ for Planck). Note that for the optimal parameter combination that can be most tightly probed by cluster counts, the relative error is even smaller ($0.5\%-0.8\%$).
 
 This implies that future cluster catalogs will allow us to probe even a tiny contribution of interacting DR and are thus a sensitive probe of dark sector interactions, independently of whether the $S_8$ tension will be confirmed or excluded with future measurements. 

\begin{table*}
\caption{\label{tab:results}%
Parameter values at $68\%$ CL intervals or $95\%$ CL upper limits for Planck 2018 TT-TE-EE data as well as forecasted CMB-S4 (tSZE clusters)$\times$ngWL (WL masses) and SPT-3G (tSZE clusters)$\times$ngWL(WL masses) datasets, assuming the ${\it Mock}_{\,\rm \Lambda CDM}$ or ${\it Mock}_{\,\rm IDM-DR}$ benchmark models.}
\scriptsize
\begin{ruledtabular}
\renewcommand{\arraystretch}{1.5}
\begin{tabular}{l|cc|cc|cc|cc|cc}
      & \multicolumn{2}{c|}{} & \multicolumn{4}{c|}{${\it Mock}_{\,\rm \Lambda CDM}$} &  \multicolumn{4}{c}{${\it Mock}_{\,\rm IDM-DR}$}\\ \hline
     & \multicolumn{2}{c|}{Planck} & \multicolumn{2}{c|}{ CMB-S4$\times$ngWL} &  \multicolumn{2}{c|}{ SPT-3G$\times$ngWL} & \multicolumn{2}{c|}{CMB-S4$\times$ngWL} &  \multicolumn{2}{c}{ SPT-3G$\times$ngWL}\\ 
      & $68\%$ limit & relative error & $68\%$ limit & relative error &$68\%$ limit & relative error & $68\%$ limit & relative error & $68\%$ limit & relative error\\ \hline
     $\xi_{\rm DR}$ & $< 0.13$ & - & $< 0.08$ & - & $<0.087$ & - & $0.119^{+0.015}_{-0.018}$ & $13.8\%$ & $0.12\pm 0.05$ & $42\%$\\ 
     $\Omega_{\rm m}$ & $0.315\pm 0.008$ & 2.5 \% & $0.316^{+0.008}_{-0.004}$ & $1.8\%$ & $0.321^{+0.012}_{-0.018}$ & $4.7\%$ & $0.309^{+0.011}_{-0.008}$ & $3.1\%$ & $0.312^{+0.024}_{-0.008}$ & $10\%$\\
     $S_8$ & $0.827\pm0.021$ & 2.4\% & $0.838^{+0.005}_{-0.007}$ & $0.7\%$ & $0.837^{+0.019}_{-0.009}$ & $1.7\%$ & $0.780^{+0.009}_{-0.007}$ & $1.03\%$ & $0.791^{+0.011}_{-0.021}$ & $2\%$ \\
\end{tabular}
\end{ruledtabular}
\end{table*}

\subsection{Effect of neutrinos}
\label{sec:neutrinos}

Massive neutrinos also have a suppression effect on structure formation and lead to a reduction of power on small scales in the matter power spectrum. Therefore, we investigate the interplay of massive neutrinos with the suppression caused by the IDM--DR interaction.  
To evaluate if we are able to distinguish between the two effects, we perform an analysis where we add the sum of the neutrino masses $\sum m_{\nu}$ as a free parameter. We focus on the combination of CMB-S4$\times$ngWL with Planck temperature and polarization data, because cluster data alone have been found to provide only a weak sensitivity to neutrino masses (see for example \cite{SPT:2018njh}).

\begin{figure}
    \begin{center}
    \includegraphics[width=0.98\columnwidth]{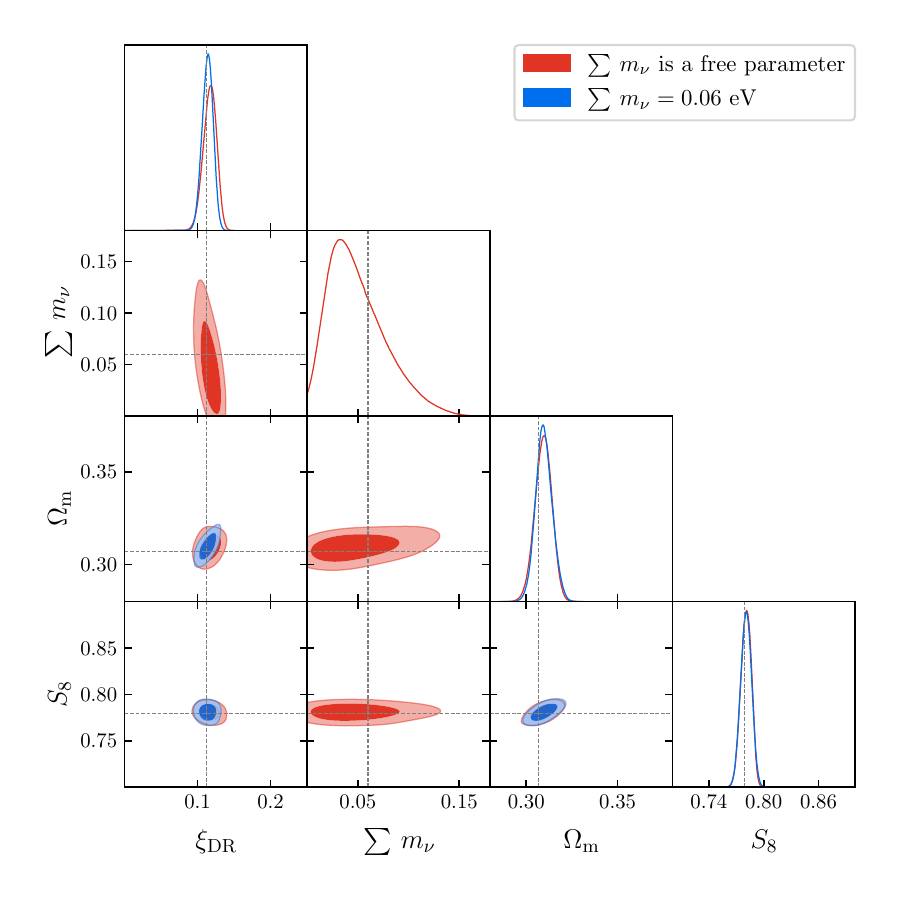}
    \caption{Comparison between the constraints from CMB-S4$\times$ngWL+Planck for the ${\it Mock}_{\,\rm IDM-DR}$ benchmark when fixing the sum of the mass of neutrinos to $0.06\,{\rm eV}$ (in blue) and adding it as a free parameter (in red). The dotted lines refer to the ``true'' values of the benchmark model.}
    \label{fig:cosmo24_nu}
    \end{center}
\end{figure}

The results for this analysis are shown in Fig.~\ref{fig:cosmo24_nu}. We find that the effect of neutrinos is distinguishable from that of IDM--DR, and the input value of $\xi_{\rm DR}$ is well recovered with a slightly larger posterior. Fig.~\ref{fig:cosmo24_nu} shows also that the degeneracy between $\sum m_{\nu}$ and $\xi_{\rm DR}$ is very weak. The projected upper bound on $\sum m_{\nu}$ of $0.103\,{\rm eV}$ 95\% CL arises primarily from CMB data.

This result shows that cluster abundance measurements combined with CMB temperature and polarization data are able to distinguish between the effect of massive neutrinos and that of dark sector interactions within the IDM--DR model. This can be traced to the following reasons: first, the suppression for each case sets in at a different scale. For neutrinos, the effect is related to their free-streaming length, corresponding to $k \sim 10^{-2}\, [h\,{\rm Mpc}^{-1}]$ for viable neutrino masses. In contrast, for IDM--DR, the suppression occurs at different scales depending on the value of $\xi_{\rm DR}$ (as described in Sec.~\ref{sec:effect_on_MPS} and demonstrated in Fig.~\ref{fig:effect_Pk}). For the specific benchmark point ${\it Mock}_{\,\rm IDM-DR}$ the suppression scale is $k \sim 10^{-1}\, [h\,{\rm Mpc}^{-1}]$. Another reason is related to having different effects on the CMB. Neutrino masses change the background evolution by contributing to the matter density at $z\lesssim 100$ while behaving as radiation at earlier times, leading to a distinct effect on the CMB anisotropy spectrum. In contrast, IDM--DR with viable values of $\xi_{\rm DR}$ leads to modifications at the level of perturbations only. Moreover, the power suppression due to IDM--DR is mostly generated already before recombination, while for massive neutrinos it is imprinted afterwards. That implies a different impact on the growth history and thus on the redshift-dependence of the power suppression.

\section{Conclusions}
\label{sec:conclusion}
In this work, we present forecasts for the sensitivity of future weak lensing mass calibration informed galaxy cluster abundance measurements on deviations from the cold dark matter paradigm. In particular, we consider an interacting dark sector featuring a dark radiation component that interacts with dark matter, similar to the well-known baryon-photon interaction.
This IDM--DR model can be realized on the microscopic level by an interaction arising from an unbroken non-Abelian gauge symmetry within the dark sector, analogous to the interactions within the Standard Model of particle physics. The non-Abelian nature of the interaction leads to a particular time-dependence of the interaction rate that imprints a suppression on the matter power spectrum that has been argued to potentially resolve the $S_8$ tension \cite{Rubira:2022xhb}.

We consider galaxy cluster samples corresponding to tSZE selected clusters from the ongoing SPT-3G as well as the future CMB-S4 mm-wave surveys, combined with cluster mass calibration constraints inferred from next generation weak lensing surveys ngWL like Euclid and Rubin. To test whether galaxy clusters will be able to differentiate between $\Lambda$CDM and IDM--DR, we have generated mock data for two different benchmark points. One based on an IDM--DR model with a lower value of $S_8$, the second based on a $\Lambda$CDM model with Planck 2018 values. The IDM--DR benchmark is chosen to be compatible with current CMB and galaxy clustering data, while at the same time supporting a lower $S_8$ value arising from dark sector interaction.

We find that an analyses of both CMB-S4$\times$ngWL and SPT-3G$\times$ngWL datasets will enable the discrimination between the IDM--DR and $\Lambda$CDM benchmark models, thereby significantly improving the sensitivity to dark sector interactions as compared to Planck 2018 CMB anisotropy data. Out of the parameters characterizing the dark sector, clusters are primarily sensitive to the dark sector temperature, parameterized by its ratio to the CMB temperature $\xi_{\rm DR}$.
For the IDM--DR benchmark, the input value of $\xi_{\rm DR}$ is recovered with a precision of $14\%$ and $42\%$ for CMB-S4$\times$ngWL and SPT-3G$\times$ngWL, respectively. The size of the catalog ($\sim 24000$ clusters) and the wide overlapping region with ngWL data, which leads to better mass calibration, yield a higher statistical power for CMB-S4$\times$ngWL over SPT-3G$\times$ngWL. For comparison, Planck 2018 data cannot discriminate the IDM--DR benchmark from $\Lambda$CDM, and therefore only sets an upper limit $\xi_{\rm DR}\leq 0.13$ at 95\% CL. When assuming the $\Lambda$CDM benchmark to describe the true cosmological model, clusters will be able to improve the upper limit to $\xi_{\rm DR} \sim 0.08$ and $\xi_{\rm DR} \sim 0.09$ for CMB-S4$\times$ngWL and SPT-3G$\times$ngWL, respectively, corresponding to an improvement of the sensitivity to the abundance of interacting DR by a factor $\sim 5-10$. Moreover, the degeneracy between $S_8$ and $\xi_{\rm DR}$ toward lower values of $S_8$, which appears when fitting Planck CMB data alone, can be lifted when combining it with future galaxy cluster and weak lensing mass calibration datasets.

It is interesting to note that SPT-3G is sensitive to IDM--DR, and that the analysis with its real data could be done already in a few years, compared to CMB-S4 whose cluster samples will be available only on timescales of a decade. In all cases, as expected, these cluster and weak lensing datasets have a constraining power on the growth of structure formation at Mpc scales, or equivalently to the parameter combination $S_8$, with a precision of $\sim 0.7-2\%$.

These results indicate that we will in fact be able to distinguish between $\Lambda$CDM and IDM--DR models using galaxy clusters. Moreover, we find that clusters combined with CMB anisotropy data can differentiate between the effect of neutrino masses and that of dark sector interactions.
Although the weak lensing informed cluster abundance datasets enable constraints on the temperature of DR ($\xi_{\rm DR}$), being related to the scale at which the power suppression sets in, they are rather insensitive to the amount of suppression caused by different fractions of IDM ($f_{\rm IDM}$) or interactions strengths ($a_{\rm dark}$). For this purpose, other datasets at smaller scales (higher $k$) like Lyman-$\alpha$ forest data are needed to constrain $f_{\rm IDM}$ and/or $a_{\rm dark}$.

\vspace*{2em}
\acknowledgments
We thank Joe Zuntz for his help with CosmoSIS. We acknowledge support by the Excellence Cluster ORIGINS, which is funded by the Deutsche Forschungsgemeinschaft (DFG, German Research
Foundation) under Germany’s Excellence Strategy - EXC-2094 - 390783311.  We acknowledge support from the Max Planck Society Faculty Fellowship program at MPE and the Ludwig-Maximilians-Universit\"at in Munich. The MCMC analysis was carried out at the Computational Center for Particle and Astrophysics (C2PAP) which is a computing facility from ORIGINS.

\appendix

\begin{figure}
    \centering
    \includegraphics[width=0.98\linewidth]{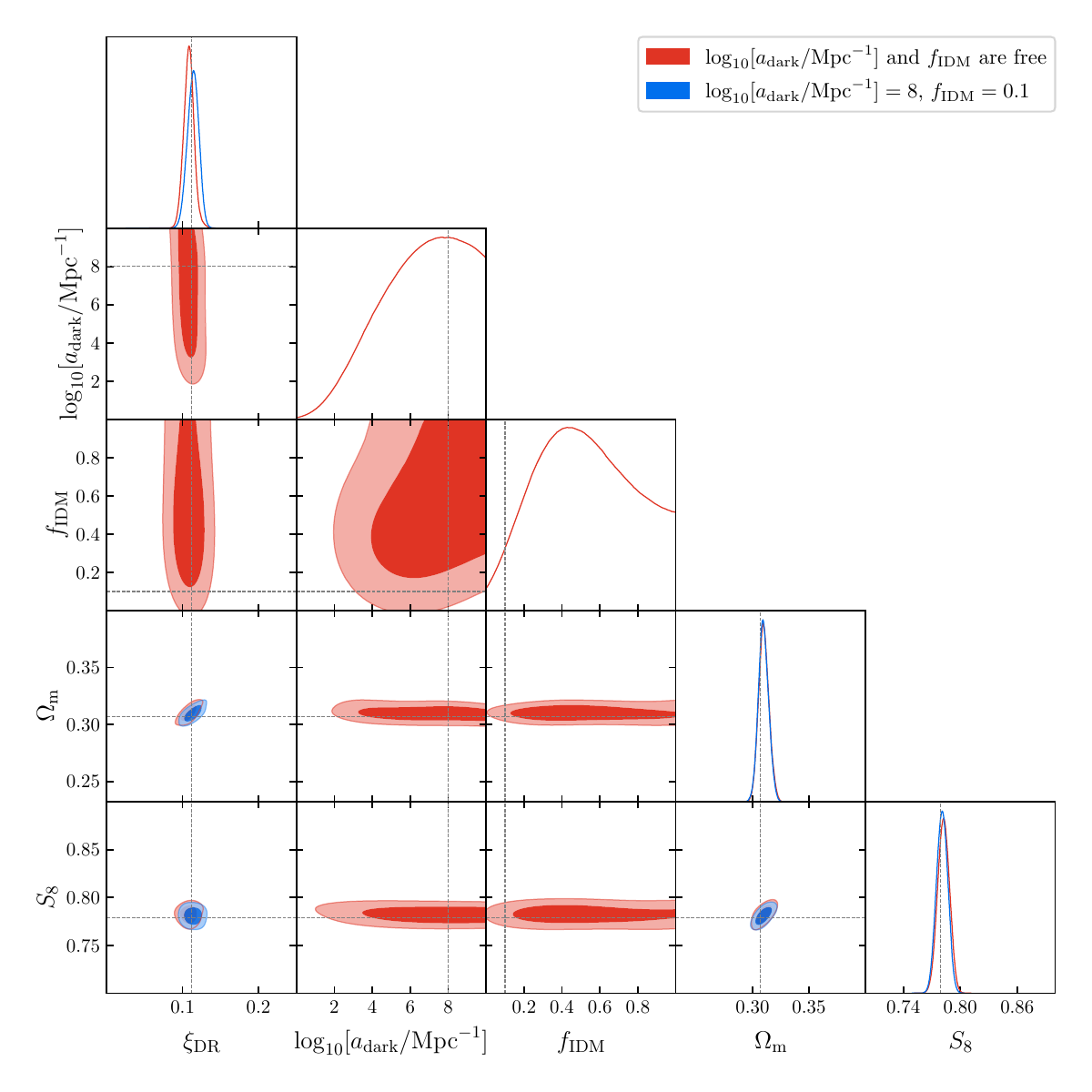}
    \caption{MCMC analysis of Planck+CMB-S4$\times$ngWL in the case of IDM--DR mock data and allowing ${\rm ln}_{10}\,[a_{\rm dark}/{\rm Mpc}^{-1}]$ and $f_{\rm IDM}$ to vary.}
    \label{fig:IDM_f}
\end{figure}

\section{Varying $a_{\rm dark}$ and $f_{\rm IDM}$}
\label{app:f_idm}

In this appendix, we show results when allowing the fraction of interacting dark matter $f_{\rm IDM}$ and the intensity of IDM--DR interaction $a_{\rm dark}$ to vary together with the other cosmological parameters. As previously discussed (see Fig.~\ref{fig:effect_Pk}), the HMF is mostly sensitive to $\xi_{\rm DR}$ and cannot constrain the other two ETHOS parameters. 
Because the amount of DR sets the matter power spectrum suppression scale and the IDM fraction is responsible for the amount of the suppression (see Sec.~\ref{sec:effect_on_MPS}), these two parameters are independent of each other and there is no degeneracy between them. Fixing $f_{\rm IDM}$ and $a_{\rm dark}$ in the analysis is attractive, because it speeds up the MCMC runs.

Nevertheless, in this appendix we take a conservative approach to test the effect of fixing $f_{\rm IDM}$ and $a_{\rm dark}$. We add $f_{\rm IDM}$ and ${\rm log}_{10}\,[a_{\rm dark}/{\rm Mpc}^{-1}]$ as free parameters to the analysis with flat priors of $\mathcal{U}(0.001, 1.0)$ and $\mathcal{U}(0.0, 10.0)$, respectively.
As shown in Fig.~\ref{fig:IDM_f}, the input value of $\xi_{\rm DR}$ is recovered and no degeneracy appears to be present with $f_{\rm IDM}$ and $a_{\rm dark}$. The parameter $f_{\rm IDM}$ is not constrained at all and takes the whole range of the prior. On the other hand, we see that clusters are able to put a lower bound on ${\rm log}_{10}\,[a_{\rm dark}/{\rm Mpc}^{-1}]$ of $\sim 1$. 

In conclusion, while the weak lensing mass calibration informed cluster abundance is very sensitive to the dark sector temperature parameter $\xi_{\rm DR}$, it is not sensitive enough to small scales ($k\gtrsim \, 0.5\,h\, {\rm Mpc}^{-1}$) to enable constraints on the other two ETHOS parameters $f_{\rm IDM}$ and $a_{\rm dark}$.  Other datasets that are sensitive to smaller scales are needed for that purpose.

\begin{figure*}
  \centering   
  \begin{overpic}[scale=0.48]{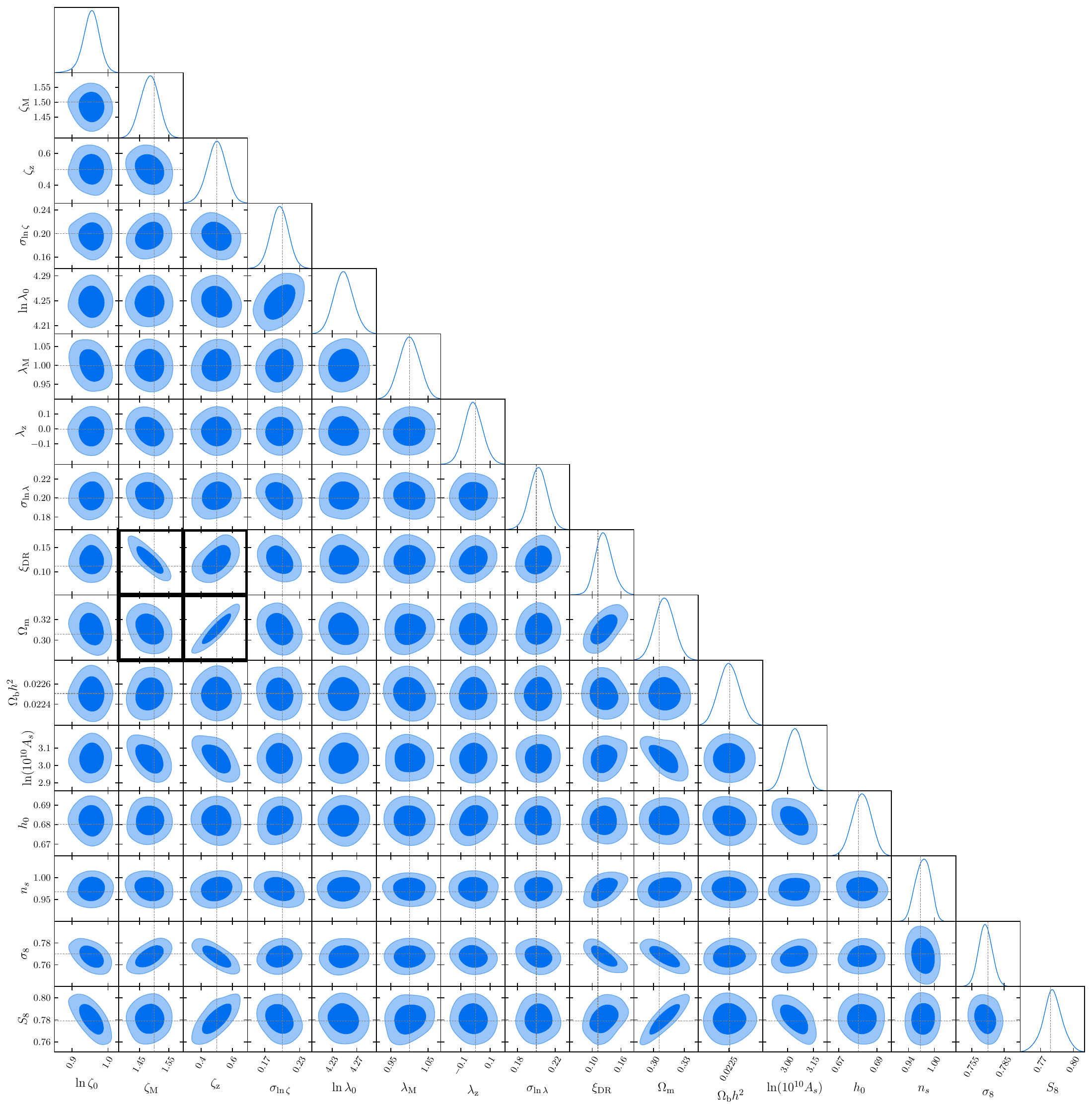}
     \put(55,60){\includegraphics[scale=0.45]{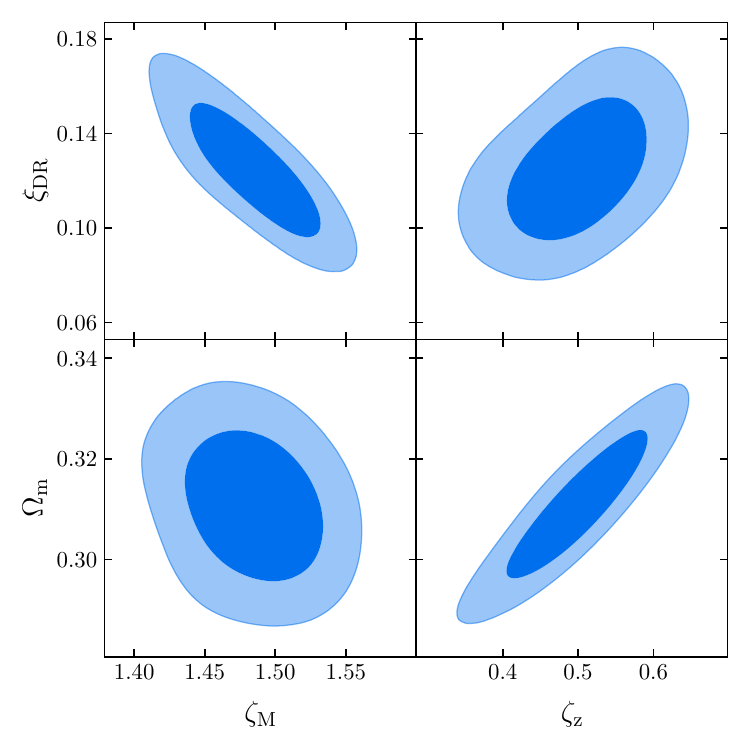}}  
  \end{overpic}
\caption{Posterior distributions of all parameters used in the analysis of CMB-S4$\times$ngWL for the benchmark model ${\it Mock}_{\,\rm IDM-DR}$.  The figure inset highlights degeneracies between tSZE observable--mass scaling relation parameters $\bsz$ and $\csz$ and the cosmological parameters $\Omega_{\rm m}$ and $\xi_{\rm DR}$ (further discussion in text).}
\label{fig:cosmo24_all}
\end{figure*}

\section{Posteriors of all parameters}
\label{app:cosmo24_all}
We show here (see Fig.~\ref{fig:cosmo24_all}) marginalized posterior distributions for all parameters considered in the analysis of CMB-S4$\times$ngWL for ${\it Mock}_{\,\rm IDM-DR}$, with the priors listed in Tables \ref{tab:scaling_relations} and \ref{tab:params}. As shown in the subplot on the right of Fig.~\ref{fig:cosmo24_all}, there is a degeneracy between some cosmological parameters and observable-mass scaling relation parameters, particularly those describing the mass- and redshift-dependence of the tSZE cluster detection significance $\zeta$. A first degeneracy we observe occurs between $\xi_{\rm DR}$ and $\bsz$. This can be traced back to the fact that higher values of the dark sector temperature ratio cause a suppression on the HMF for more massive halos (see Fig.~\ref{fig:effect_Pk}), which is then compensated by a higher value of $\bsz$ (see Eq.~\ref{eq:zetaM}). This degeneracy could be mitigated by improving the constraint on $\bsz$ by using a sample with a broader range of masses. This could for example be realized by adding low-mass clusters detected via X-ray emission. There is a weaker degeneracy between $\xi_{\rm DR}$ and $\csz$ that can be understood in a similar way.
A much stronger degeneracy occurs between $\csz$ and $\Omega_{\rm m}$. A way to better constrain $\csz$, thereby improving $\Omega_{\rm m}$ and $\xi_{\rm DR}$ constraints, is to increase the redshift range of the cluster and weak lensing datasets. Increasing the low redshift cluster sample could help. In addition, CMB lensing data hold great promise for constraining the masses of the highest redshift clusters, where the ground based ngWL weak lensing samples lose much of their sensitivity.

\newpage
%

\end{document}